\documentclass[preprint,preprintnumbers,amsmath,amssymb,superscriptaddress,showpacs,showkeys]{revtex4}

\usepackage{graphicx}% Include figure files
\usepackage{bm}% bold math

\begin{document}

\texttt{preprint}

\title{Introduction to Graphene Electronics -- A New Era of Digital Transistors and Devices}

\author{K.C. Yung}
\affiliation{
Department of Industrial and Systems Engineering, The Hong Kong Polytechnic University, Hung Hom, Kowloon, Hong Kong, China}
\author{W.M. Wu}
\affiliation{
Department of Physics, Loughborough University, Loughborough LE11 3TU,
United Kingdom}
\author{M.P. Pierpoint}
\affiliation{
Department of Physics, Loughborough University, Loughborough LE11 3TU,
United Kingdom}
\author{F.V. Kusmartsev}
\email{f.kusmartsev@lboro.ac.uk}
\affiliation{
Department of Physics, Loughborough University, Loughborough LE11 3TU,
United Kingdom}

This is an updated version of our article, due to be published in Contemporary Physics (Sept 2013). Included are updated references, along with a few minor corrections.\\ \\ \\ \\ \\ \\ \\

\begin{abstract}
The speed of silicon-based transistors has reached an impasse in the recent decade, primarily due to scaling techniques and the short-channel effect. Conversely, graphene (a revolutionary new material possessing an atomic thickness) has been shown to exhibit a promising value for electrical conductivity. Graphene would thus appear to alleviate some of the drawbacks associated with silicon-based transistors. It is for this reason why such a material is considered one of the most prominent candidates to replace silicon within nano-scale transistors. The major crux here, is that graphene is intrinsically gapless, and yet, transistors require a band-gap pertaining to a well-defined ON/OFF logical state. Therefore, exactly as to how one would create this band-gap in graphene allotropes is an intensive area of growing research. Existing methods include nano-ribbons, bilayer and multi-layer structures, carbon nanotubes, as well as the usage of the graphene substrates. Graphene transistors can generally be classified according to two working principles. The first is that a single graphene layer, nanoribbon or carbon nanotube can act as a transistor channel, with current being transported along the horizontal axis. The second mechanism is regarded as tunneling, whether this be band-to-band on a single graphene layer, or vertically between adjacent graphene layers. The high-frequency graphene amplifier is another talking point in recent research, since it does not require a clear ON/OFF state, as with logical electronics. This paper reviews both the physical properties and manufacturing methodologies of graphene, as well as graphene-based electronic devices, transistors, and high-frequency amplifiers from past to present studies. Finally, we provide possible perspectives with regards to future developments.
\end{abstract}


\maketitle

\section{Introduction}
\begin{figure}
\begin{center}
\includegraphics[width=15cm]{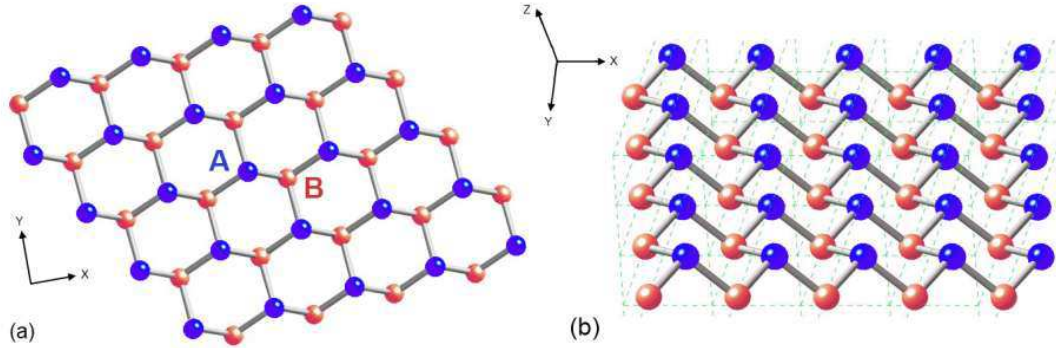}% Here is how to import EPS art
\caption{\label{honeycomb} The two-dimensional honeycomb lattice - carbon atoms A and B per unit cell are shown. The form of graphene, or other two-dimensional crystals (such as silicene, BN or germanene) can be approximately flat, or slightly corrugated (cf. side-view of Figure \ref{Fig1ab}b) \cite{O'Hare}. This corrugation depends upon the physical conditions of graphene.}
 \label{Fig1ab}
\end{center}
\end{figure}

Not long before graphene was first manufactured by the Manchester research group in 2004 \cite{Novoselov1,Geim1,Geim2,Geim3}, theorists still believed that such two-dimensional structures were unstable due to thermal fluctuations \cite{Geim3,Meyer}, famously referred to as the Landau--Peierls arguments (cf. also Mermin--Wagner theorem \cite{Meyer,O'Hare,O'Hare1}). Recently, the paradox behind graphene's existence has been resolved \cite{Meyer,O'Hare}, and that it can be stabilised by transverse lattice distortions \cite{Tapaszto}. Stable forms of various other two-dimensional crystals such as graphene, silicene and germanene have all been attained \cite{O'Hare,Drummond}. 
Graphene was the first example which is able to exist in a single atomic layer with honeycomb hierarchy \cite{Novoselov1} (cf. Figure \ref{Fig1ab}). It is composed of a single layer of carbon atoms, and can be extracted from graphite with full preservation of the hexagonal honeycomb structure (also referred to as chicken wire for quantum information processing \cite{Fal'ko}). This material has astonishing properties: it is stronger than diamond, more conductive than copper and more flexible than rubber. Graphene has primarily attracted the attention of scientific and engineering communities, due to its outstanding electrical, thermal and optical  properties \cite{Novoselov2,Neto,Chen,Rafiee}, displaying having a strong potential for many applications. Graphene has also exhibited anomalous quantum behaviour at room temperature, such as Klein tunneling, relativistic quantum Brownian motion \cite{Novoselov3,Katsnelson,Pototsky,Zalipaev} and the Quantum Hall Effect (QHE) \cite{Dean}. These phenomena have also drawn strong attention in the field of quantum engineering \cite{Trauzettel,Zagoskin}. The 2010 Nobel Prize \cite{Geim2,Neto} is not the end of the story, it is merely the beginning of an adventure - how to make graphene applicable, being our main concern. Graphene sensors, transistors, and a number of electronic nano-devices are emerging in laboratories every day \cite{Iyechika}. Furthermore, graphene is an easily manufactured material - electrically compatible with many existing key materials, such as silicon.
	
Graphene is the thinnest material in the world - a single layer of carbon atoms with a hexagonal structure \cite{Novoselov1} (cf. Figure~\ref{honeycomb}a). It is one of the allotropes of carbon, diamond being another example, albeit with different physical properties. Graphene layers stack together to form a graphite crystal lattice, and can be found in the form of highly ordered pyrolytic graphite (HOPG) or even the core of a pencil \cite{Savage}. There are four electrons present in the outer shell of a single carbon atom. For graphene, there are three $\sigma$ bonds and one $\pi$ bond present. The $\sigma$ bonds are an $sp^2$ orbital hybridisation - a mixture of orbitals $(s, p_x, p_y)$, thus providing a strong binding force for its atomic neighbours. The remaining electron $p_z$ orbital constitutes the $\pi$ bond. These are responsible for the half-filled band that enables free-moving electrons, thus displaying a metallic characteristic \cite{Wallace}. Graphene is also capable of being combined with hydrogen or oxygen to form `graphane', or other carbonic compounds respectively.

At low temperature, suspended graphene may also give rise to small quantum corrugations, whereby the two sub-lattices (A and B) are shifted with respect to each other in a transverse direction (cf. Figure~\ref{honeycomb}b). The distance is of the order equivalent to some fraction of the inter-atomic spacing. At room temperature, the shape is nearly flat, with corrugation remaining only on the large scale \cite{Meyer}. Here, low temperature microscopic quantum corrugations are transformed into transverse thermal vibrations.
	
In 1947, Wallace \cite{Wallace} first studied the electronic band structure of monolayer graphite (graphene) by applying the simplest tight-binding model with a single hopping integral. He showed that the conduction band is half-filled, with an electronic spectrum that is gapless at six points of the Brillouin Zone. In vicinity of these points, the energy-momentum spectrum has a conical, Dirac form. These six points are therefore referred to as the Dirac points. He concluded that within two-dimensional graphene, an electrical conductivity should theoretically exist. A year later, Ruess and Vogt \cite{Geim2,Dreyer,Ruess} observed thin-film graphene oxide (GO) pieces using transmission electron microscopy (TEM). A single layer of graphene oxide was then discovered by Boehm {\it et al.} \cite{Geim2,Boehm1,Boehm2} in the early 1960s, and it was Boehm {\it et al.} who first proposed the name `graphene' in their studies \cite{Boehm1,Boehm2}. Nevertheless, all these systems were found to be insulating. Although research on graphene, graphene-like structures and epitaxial graphene films continued for decades (cf. for example \cite{Gall,Berger,Sadowski}), it has significantly intensified since 2004, when Geim and his fellow researchers \cite{Novoselov1} successfully segregated graphene flakes from graphite using the simple scotch-tape method \cite{Geim1,Novoselov4}. This process will be elaborated upon in Section II.A. They subsequently went on to show that a graphene layer, separated from graphite, is an example of a two-dimensional structure possessing outstanding electrical properties \cite{Novoselov1,Novoselov4}. It is therefore no surprise that a huge number of studies on the physical and chemical properties of graphene have emerged in recent decades.

\begin{figure}
\begin{center}
\includegraphics[width=8cm]{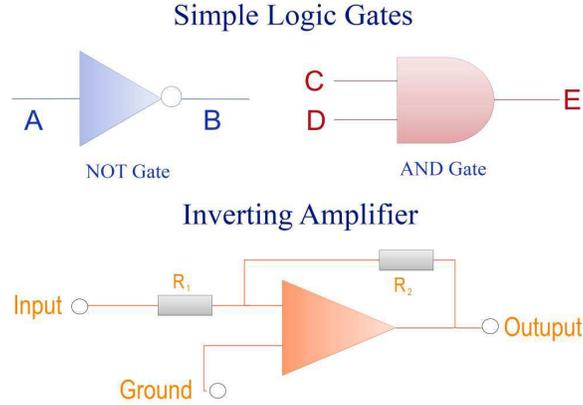}% Here is how to import EPS art
\caption{\label{logic} Logic gates are the fundamental building blocks of electronic devices -- the transistor being a key example. Here, the diagram shows the basic `NOT' and `AND' logic gates. The input/output signals are generally either one or zero. The inverting amplifier is also shown. The underlying principle is dependent upon the ratio of the resistors $R_1$ and $R_2$, with an output voltage $V_{out} = - (R_2/R_1)\,V_{in} $. This output voltage can be adjusted by the varying the resistors in the circuit.}
\end{center}
\end{figure}

Recently, there has been strong progress in development of many graphene applications. Graphene's unique electrical properties make it an ideal material for many electronic devices -- in particular, graphene based transistors. In the present article, we describe various designs of graphene transistor that have been recently proposed, and how they capitalise upon graphene's high electron mobility and conductivity.

The field-effect transistor (FET) was first reported by J. Barden and W. H. Brattain in the 1940s at Bell Laboratories \cite{Bardeen}. A transistor is the basic building block of many electronic circuits, normally acting as a logic gate or inverting amplifier in integrated circuits (cf. Figure~\ref{logic}). According to Moore's law, the size of a circuit decreases roughly by half, every 1.5 years \cite{Schwierz}. However, the scaling techniques, as well as the speed of the devices, have both reached a bottleneck in recent years. This is mainly because the decreasing scale leads to an increase of the energy dissipation per unit area (i.e., power density). The operating speed is also limited by the mobility and thermal conductivity of existing materials. New materials are thus highly sought after, to alleviate such aforementioned drawbacks. This article gives a brief review of the most up-to-date properties of graphene, electronic devices and transistors, together with their operational mechanism.

\section{Existing Fabrication Methodologies}
The type of graphene (characterised by different properties) depends strongly upon its method of preparation. Three methods in particular, include mechanical exfoliation from bulk graphite \cite{Novoselov4}, graphitisation of silicon carbide (SiC) substrates (epitaxial growth) \cite{Heer,Emtsev}, and chemical vapor deposition (CVD) on various transition metals \cite{Chen2,Kim,Li}. In terms of the quality of graphene (such as mobility or structural defects), the exfoliated graphene produces the best material, followed by epitaxial graphene, and finally, CVD methods.

\begin{figure}
\begin{center}
\includegraphics[width=8cm]{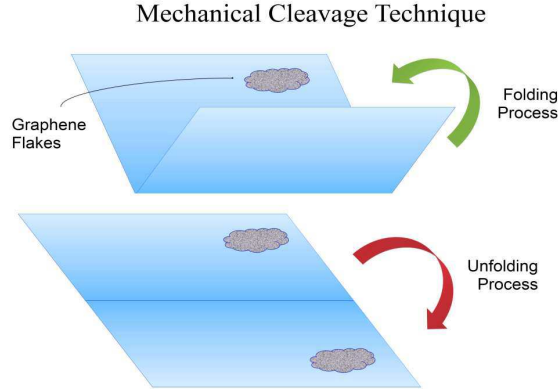}% Here is how to import EPS art
\caption{\label{mech} The simplest method of segregating graphene from graphite, via the `scotch-tape' method. By placing graphite flakes upon the tape, and then repeating the folding/unfolding process, we can thus obtain one or few layers of graphene. Simple optical instrumentation (via the principle of light interference) can be used to verify the existence of graphene.}
\end{center}
\end{figure}	

\subsection{Mechanical Cleavage}
Mechanical cleavage is one of the simplest methods of fabricating a graphene layer \cite{Geim1,Geim2,Geim3,Katsnelson,Novoselov4}. The process is similar to a folding and unfolding technique (as shown in Figure~\ref{mech}). One has to first cut the surface of a graphite sample \cite{Geim1,Geim2,Geim3}, usually with highly ordered pyrolytic graphite (HOPG) being used \cite{Iyechika}. A small piece of graphite is then placed upon the sticky surface of adhesive tape \cite{Iyechika}. The tape is folded and unfolded, with gradually more and more layers being stripped away \cite{Geim2}. The graphite is therefore made thinner - the procedure being repeated until what remains is one, or few layers of graphene. Monolayer graphene can be verified through a simple optical microscope via light interference patterns \cite{Novoselov4}, or also via Raman spectroscopy \cite{Geim3}. The mobility of exfoliated graphene can reach 15,000\ cm$^{2}$V$^{-1}$s$^{-1}$ \cite{Schwierz}. Despite being a relatively simple way to produce graphene, this method produces only small quantities each time \cite{Novoselov4}. 

\subsection{Vacuum Epitaxial Growth Technique}
An alternative means of fabricating graphene is via the epitaxial growth technique. Graphene produced in this way would usually possess an intermediate range of values for the mobility, smaller than for exfoliated graphene, but larger than for CVD graphene. Silicon carbide (SiC) is coated on a silicon wafer, with high temperatures being required \cite{Geim1,Geim3,Iyechika}. Both silicon and carbon atoms would then individually become segregated. As the silicon atoms are evaporated by temperatures in excess of $1,000{\rm K}$, only carbon atoms remain \cite{Geim1,Geim3,Iyechika} and graphene will subsequently grow on the substrate \cite{Iyechika}. Regardless, structural defects are always found since carbon atoms will sometimes be burnt due to the high temperatures - thus being contaminated by oxygen and hydrogen \cite{Iyechika}. 

The quality of epitaxial graphene can be conveniently analysed with the use of Raman spectroscopy. In particular,  studies of the Raman active, single phonon (G band) and the two-phonon (2D band) are popular \cite{Hlawacek,Robinson,Trabelsi}. Analysis of the position and shape of these Raman scattering lines provide much useful information. For example, the observation of a high-frequency shift of the G peak may describe graphene doping and associated changes in electron mobility \cite{Das}.  It is also highly beneficial for the monitoring of dopants (via Raman scattering) in an electrochemically top-gated graphene transistor. The local mapping of these Raman modes on the graphene surface (a method known as Raman topography) may describe the charge puddles created in graphene, as well as many other inhomogenous structures.    

In particular, recent studies of the Raman topography of epitaxial graphene, show that the electron mobility is highly dependent upon thickness and monolayer strain uniformity \cite{Hlawacek,Robinson,Trabelsi}. At room temperature, the highest mobility of epitaxial graphene (on the carbon-terminated face of the SiC substrate) reaches a value of 5,000\ cm$^{2}$V$^{-1}$s$^{-1}$ \cite{Schwierz}. It was also shown that the carrier mobility depends strongly upon the stacking of graphene layers \cite{Hlawacek}. 
It is worth mentioning that for exfoliated graphene (produced by mechanical cleavage method (cf. section II.A)), the mobility is higher when compared to that via the epitaxial growth technique \cite{Novoselov1,Schwierz}. 

\begin{figure}
\begin{center}
\includegraphics[width=8cm]{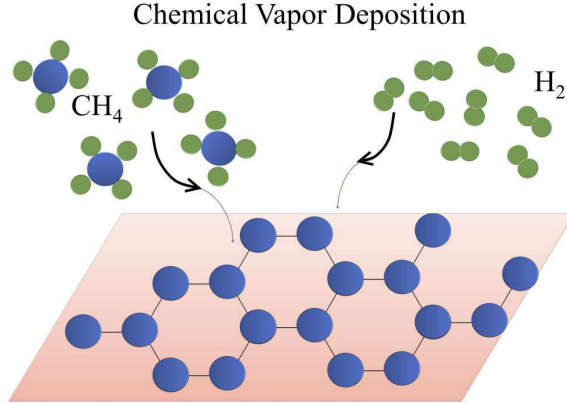}% Here is how to import EPS art
\caption{\label{CVD} Chemical vapor deposition (CVD) is a high-temperature chemical process, subjecting catalytic metals to hydrocarbon gases. Gases such as  hydrogen ($H_2$) and methane ($CH_4$), are both added to the chamber. The high temperatures decompose the molecular bonds, until all hydrogen has been burnt out. What remains is a single layer of carbon atoms being deposited on the metal surface.}
\end{center}
\end{figure}

\subsection{Chemical Vapor Deposition}
Chemical vapor deposition (CVD) is usually used in industry \cite{Iyechika}, since it can be easily mass produced. The process involves high-temperature chemical reactions with hydrocarbon gases (e.g. $H_2$ and $CH_4$) and a catalytic metal surface such as either copper or nickel (cf. Figure 4). These metals help decompose the chemical bonding \cite{Geim1,Geim3,Iyechika}, and order carbon atoms into a hexagonal (honeycomb) lattice. Hence, the larger the metal surface, the larger the quantity of graphene that can be produced. For this purpose, the most popular is a polished polycrystalline copper foil. It is easy to transfer graphene from this metal surface to other substrates -- for example, by applying a voltage. Therefore, the CVD method exhibits large scale superiority with respect to other graphene production methods. It is a relatively simple and low-cost method which allows the growing of monolayer graphene to large size \cite{Bae}. However, the charge carrier mobility of CVD graphene is generally lower than data reported for exfoliated graphene \cite{Bae,Du,Chen3,Bolotin,Morozov,Kusmartsev}. It was found that the main source of this low mobility is the presence of grain boundaries \cite{Song}. This has been well studied with the use of both transmission electron microscopy (TEM) and scanning electron microscopy (SEM). The point defects, as well as surface contamination, are further key factors which significantly affect the mobility. The mobility of a short channel device, where the grain boundary is absent, can reach 2,700\ cm$^{2}$V$^{-1}$s$^{-1}$. In the presence of this boundary, it will be roughly three or four times smaller \cite{Song,Hlawacek}.

\section{Physical Properties of Graphene}
\subsection{Graphene Electrical Conductivity}
Mobility is a measure of how fast the carriers propagate in an electric field. Its value depends upon the type of material and the interaction with substrates. After the process of annealing, some suggest the mobility of exfoliated graphene can reach values upwards of 200,000\ cm$^{2}$V$^{-1}$s$^{-1}$ \cite{Neto,Schwierz,Kusmartsev} in a perfect structure. Recently, experiments with graphene \cite{Sarma} show that at this specific point, there should be a minimum conductivity in graphene $\sigma_{0} \sim {4e^2}/{h}$ \cite{Geim3}. This is around double the value for the conductance quantum $2e^2/h$, observed in quantum wires \cite{Yang-Bo}. This value is independent of the material properties and length of the quantum wire \cite{Datta}.
It is also of particular interest that the concentration of electrical carriers at the Dirac point tends to zero. Here, even at room temperature, the charge-carriers may perform long range (ballistic) transport without much scattering at all \cite{Geim3,Du,Kusmartsev}. Graphene is also useful for electronics, especially due to its elasticity and strong resistance against destruction. Young's modulus for graphene can be as large as 1TPa \cite{Neto}. Graphene is the strongest material in the world, around 100 times stronger than steel \cite{Savage}.  Yet, due to its monolayer structure (cf. Figure \ref{Fig1ab}), it can be stretched and bent rather easily.

\subsection{Thermal Conductivity}
Using high-resolution vacuum scanning thermal microscopy, it was found by Pumarol {\it et al.} \cite{Pumarol} that thermal conductivity in suspended graphene is carried by ballistic phonons. Hence, the thermal conductance of graphene on a substrate will significantly downgrade, since the number of scattering channels increases. In this case, heat may be lost to the substrate. The thermal conductance also decreases with an increased number of graphene layers. Analysing nano-thermal images, it was found that the mean-free-path of thermal phonons in graphene on a substrate, is smaller than 100nm. According to Prasher \cite{Prasher}, the thermal conductivity of graphene on a silicon-dioxide substrate can reach around 600\ Wm$^{-1}$K$^{-1}$, incidentally higher than copper. Other papers provide even higher values for the in-plane thermal conductivity, but of the same order of 1,000\ Wm$^{-1}$K$^{-1}$ \cite{Balandin,Seol}. The heat flow between graphene and the substrate, would influence the dissipation, due to the high in-plane thermal conductivity \cite{Koh}. Thus, in-plane heat transport for graphene is associated with acoustic phonons \cite{Pumarol,Seol}. However, the mechanism of heat transport across graphene interfaces in the cross-plane direction is currently unknown, as mentioned by Koh {\it et al.} \cite{Koh}

\subsection{Photonic properties}
Single layer graphene absorbs a mere $\pi\alpha \approx 2.3\%$ of visible light (where $\alpha = 1/137$ is the fine structure constant \cite{Nair}), thus possessing a high transparency \cite{Rafiee}. Graphene provides a wide range of absorption across the visible light spectrum. Bonaccorso {\it et al.} \cite{Bonaccorso} mention that the peak of absorption is in the ultraviolet domain, because of the van Hove singularity. The peak arises due to the van Hove singularity in the energy density of states (DoS). It is, in fact, a well suited material for photonic devices. Being almost perfectly transparent to visible light due to its atomic thickness, such a material can be made available for small screen devices \cite{Rafiee,Savage,Prasher}. For similar reasons, there is also common belief that graphene's true potential lies in opto-electronics, as discussed by Nair {\it et al.} \cite{Nair} and Bonaccorso {\it et al.} \cite{Bonaccorso}. 

Stoehr {\it et al.} \cite{Stoehr} observed white optical luminescence of graphene, when subject to picosecond infrared laser pulses - this creates a high density electron-hole plasma. The luminescence arises due to electron-hole recombination, created in a broad energy range. There are two relaxation mechanisms observed in the graphene electron-hole plasma \cite{Breusing}. The first relaxation mechanism relates to the intra-band electrons and holes collisions, with a duration of around hundred femtosecond. The second one arises from the cooling of hot optical phonons, with around a few picoseconds, which is slower than first mechanism \cite{O'Hare,Bonaccorso,Kampfrath,Lazzeri,O'Hare2}.

\begin{figure}
\begin{center}
\includegraphics[width=10cm]{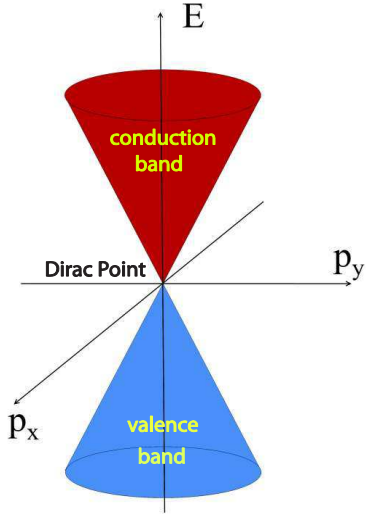}% Here is how to import EPS art
\caption{\label{cone} In graphene, the conduction and valence bands touch at six points of the Brillouin Zone. These are referred to as the Dirac points. In the vicinity of these points, both the conduction (red) and valence (blue) bands have a conical shape. These cones can be used to describe the low-energy-momentum dispersion relationship, $E=\pm v_{F} \sqrt{p_x^2+p_y^2}$, where $v_{F}$ is the Fermi velocity and  $(p_x,p_y)$ defines the electron momentum. The Fermi energy level $E_F$ usually coincides with the Dirac point - where the density of states (DoS) is equal to zero.  Electrons occupy all states in the valence band, with the conduction band remaining unoccupied.}
\end{center}
\end{figure}	

\subsection{Energy Spectrum and Anomalous Quantum Hall Effect}
Graphene has a closely packed, periodic structure of carbon atoms, and consists of two carbon sub-lattices (A and B) per unit cell \cite{Savage} (cf. Figure~\ref{honeycomb}a). As previously mentioned, there are six Dirac points in the corners of Brillouin zone, with a linear, conical relationship being found in the lower part of the energy-momentum spectrum. This can be approximated by the following dispersion relation \cite{Wallace}, 
\begin{equation}
 E(p_x,p_y)=\pm v_{F} \sqrt{p_x^2+p_y^2},
 \label{D-spectrum}
\end{equation}
where $v_{F}$ is Fermi velocity, and $p_x$ and $p_y$ are the electron momenta in the graphene plane. The points in momentum space where the energy vanishes (i.e., $E(p_x,p_y)=0$) are called the Dirac points.
A single Dirac cone for the graphene energy spectrum is shown in Figure~\ref{cone}. At the Dirac point, there is a direct contact of the conduction and valence bands \cite{Novoselov1,Novoselov2,Katsnelson}, possessing a zero energy band-gap \cite{Geim3,Iyechika,Wallace}. Graphene also exhibits zero effective mass, and is well described by the relativistic Dirac Hamiltonian and equation:
\begin{equation}
\hat{H} = v_{F} {\hat \sigma} \cdot {\bf p} ,\ \ \ \ \ \ \ \ \ \  \hat{H}\psi=E\psi
\label{Dirac-Ham}
\end{equation}
%or the Dirac equation $\hat{H}\psi=E\psi$, 
where $ {\hat \sigma} \cdot {\bf p}= \sigma_x p_x+\sigma_y p_y$, and  $\psi = \begin{pmatrix} \psi_A \\ \psi_B \end{pmatrix}$. Electron momentum is given by ${\bf p}=(p_x,p_y)$ \cite{Geim3}.
Wave functions $\psi_A$ and $\psi_B$ relate to sub-lattices A and B respectively, and  ${\hat \sigma}$ refers to the vector of $2 \times 2$ Pauli matrices ${\hat \sigma}=(\sigma_x,\sigma_y)$: 
\begin{equation} 
\sigma_x = \begin{pmatrix}
0 & & 1 \\
1 & & 0
 \end{pmatrix},\ \ \ \ \ \ \ \ \ \ \ \ \sigma_y = \begin{pmatrix}
0 & & -i \\
i & & 0
 \end{pmatrix}
 \end{equation}
The eigenvalues of this 2$\times$2 Hamiltonian $\hat{H}$, immediately give the Dirac spectrum (cf. equation (\ref{D-spectrum})). Furthermore, charge-carriers in graphene travel at 1/300 the speed of light \cite{Geim3,Fal'ko,Katsnelson} ($v_{F} \approx 10^6\ {\rm ms}^{-1}$), thus providing a miniaturised platform to test the phenomena of quantum electrodynamics (QED) \cite{Geim3} without the use of expensive accelerators. In general, quantum behavior is normally exhibited at low temperature; such processes include the QHE, where the transverse Hall resistance is quantised. However, with graphene, the possibility has now arisen to study such quantum behavior at room temperature \cite{Novoselov3,Katsnelson}. Some of the most recent experimental research \cite{Novoselov2,Novoselov3,Dean} has reported observation of the QHE within a graphene monolayer, displaying quantum resistance at room temperature. Integer QHE in graphene is slightly different to that of standard two-dimensional semiconducting devices. Here, the quantised conductivity has an additional shift equal to one half of the minimum conductivity $\sigma_{0}$. The fractional QHE in graphene is associated with the electron-electron interaction as in other semiconductors, and a four-fold degenerate state originating due to the existence of the two degenerate sub-lattices (A and B) \cite{Geim3,Novoselov2,Novoselov3,Dean,Jiang}. Both fractional and anomalous QHE have been observed within the single monolayer, as well as bilayer graphene \cite{Novoselov3,Katsnelson,Dean}. 
  	
\begin{figure}
\begin{center}
\includegraphics[width=8cm]{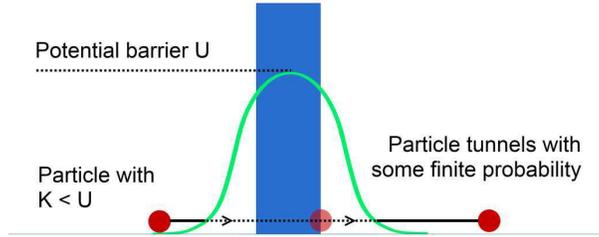}% Here is how to import EPS art
\caption{\label{klein} Klein tunneling with perfect transmission through the barrier is shown. In classical mechanics, a particle with kinetic energy $K$ is incapable of surpassing a potential barrier $U$, for the condition $K<U$. However, quantum mechanics allows a particle to tunnel through the barrier with some finite probability, even when $K<U$. This becomes most unusual in the case of graphene. For certain angles, a particle can tunnel through the barrier with perfect transmission.}
\end{center}
\end{figure}		
	
\subsection{Quantum Klein Tunneling}	
Graphene provides a new quantum feature with regards to tunneling \cite{Katsnelson2,Calogeracos1,Su,Calogeracos2}. In classical mechanics, an electron with kinetic energy $K$ cannot penetrate a potential barrier $U$ when $K < U$. Interestingly, quantum theory shows that there is a certain probability of transmission, even when $K < U$. Furthermore, under certain conditions, relativistic quantum mechanics permits an electron to tunnel in a straightforward manner through the barrier with perfect transmission \cite{Calogeracos1} - a process called Klein tunneling. Katsnelson {\it et al.} \cite{Katsnelson2} reveal that Klein tunneling can occur in graphene, as depicted in Figure \ref{klein}. It is further surprising that a perfect transmission occurs in graphene \cite{Katsnelson2}. Overall, it appears as if electrons penetrate the square potential wall, without restriction being placed upon on its height. In reality, it is very difficult to manifest a perfectly square potential -- graphene will always have a smooth potential. 

Recently, Zalipaev {\it et al.} \cite{Zalipaev} discovered that certain shapes of potential well (e.g., parabolic) would influence the tunneling behaviour of an electron in graphene (cf. for details \cite{Zalipaev}).
There is a drastic difference between tunneling through smooth and rectangular barriers. The Klein tunneling features are clearly exhibited only for a rectangular barrier. The type of the tunneling behavior depends upon the energy of the particle relative to the barrier height, and the angle at which the particle is incident to the potential barrier. When the energy of the particle is close to the top of the barrier, one observes conventional tunneling. For intermediate energies, the smooth potential acts as a Fabry-Perot interferometer. However, for energies close to the Dirac point, confined bound states will arise \cite{Zalipaev}. The confinement effect at small energies for the special waveguide geometry, can also be found in references \cite{Hartmann,Wu,Williams}. %[65-67]. 
The conducting waveguide channel in electronic devices is usually created by a gate voltage applied to graphene. This leads to the lateral confinement of electrons within the channel. However, electrostatic potentials of the circular form which are very smooth (i.e., do not have exponentially decaying tails), may possess a threshold for the appearance of the zero-energy confined state \cite{Downing,Stone}. This may lead to an unusual Aharonov-Bohm (AB) effect in graphene (e.g., the fractional AB effect arising in strongly correlated electrons) \cite{Kusmartsev-1995}.

\subsection{Quantum Capacitance}
Quantum capacitance is one of the most interesting properties of the graphene capacitor, as reported by Yu {\it et al.} \cite{Yu}. Total capactiance $C$ in graphene is given by ${1}/{C}={1}/{C_{es}}+{1}/{C_{q}}$, where $C_{es}$ is related to classical electrostatics, and $C_{q}$ is the quantum capacitance. $C_{q}$ is proportional to ${dn}/{d \mu}$ (where $n$ is the electron density and $\mu$ is the chemical potential), and also the number of electron states (spin up/down) near the K and K' points of the Brillouin zone \cite{Yu}. The density of states (DoS) in two-dimensional graphene associated with energy $E$ is given by \cite{Katsnelson}
\begin{equation}
D(E) = \frac{8 \pi | E |}{h^2 v^2_{F}},
\end{equation}
where Planck's constant $h=6.6261 \times 10^{-34}$Js, with a Fermi velocity $v_F \approx$ $10^6 \rm{ms^{-1}}$ for graphene. At zero temperature, the Fermi energy $E_F$ coincides with the chemical potential $\mu$. The quantum capacitance can then be described by the analytic expression
\begin{equation}
C_{q} = A e^2 \frac{dn}{d \mu}= \frac{A e^2 8\pi  | \mu |}{h^2 v^2_{F}},
\end{equation}
where $A$ is the surface area of the capacitor electrodes \cite{Yu} and the electron density $n$ is related to $\mu$ via the equation:
\begin{equation}
n=\int_0^{\mu} D(E) dE=  \frac{g \mu^2 \pi} {h^2 v_F^2}
\end{equation}
The degeneracy factor $g$ takes into account the double spins, and valley degeneracy of the Dirac spectrum for graphene (i.e. $g=4$).
For zero applied electric field (i.e., zero gate voltage), both the chemical potential $\mu$ and DoS are very small.  At low temperatures, the chemical potential then takes a very simple form:
\begin{equation}
\mu  =  \frac{h v_{F}} {2} \sqrt{\frac{n} {\pi}}.
\end{equation}
For example, consider an epitaxial graphene layer doped to an electron density $n= 10^{12}$ cm$^{-2}$. The chemical potential is then shifted above the Dirac point due to the doping. We can immediately estimate both the chemical potential and the Fermi energy, $\mu=E_F$ = 0.1167 eV.
Then, the quantum capacitance per unit area is equal to $C_q/A=27.46 \times 10^{-3}\ \rm{Fm^{-2}}$. 
For pure graphene subject to a small applied voltage, $C_{q}$ is very small, and hence the dominant contribution to the total capacitance \cite{Kusmartsev}. This is the so-called graphene quantum capacitance effect (cf. for details Xia {\it et al.} \cite{Xia} and Giannazzo  {\it et al.} \cite{Giannazzo}). However, as an applied electric field is increased, more electrons occupy the conduction band, and $\mu$ shifts to an upper value. $C_q$ becomes larger compared to $C_{es}$, with the quantum capacitance effect becoming less important. Recent studies have discovered that $C_{q}$ can reduce significantly due to disorder of the graphene structure, as mentioned by Li {\it et al.} \cite{WLi}. In general, the contribution $C_{es}$ is much smaller than $C_{q}$ in the presence of an electric field. Therefore, the capacitance of graphene simply reduces to the electrostatic case ($C=C_{es}/(1+C_{es}/C_q) \approx C_{es}$).

\section{Graphene Materials}
It is important to have a brief introduction of the electronic properties of carbon nanotubes (CNTs) \cite{Avouris,Tans,Zhu,Martel,McCann} and graphene nanoribbons (GNRs) \cite{Li2} before discussing graphene devices \cite{Park,He,He2}. Fal'ko and McCann first showed that both GNRs and single-wall CNTs are described by the Dirac Hamiltonian of equation (\ref{Dirac-Ham}), with various periodic and hard-wall boundary conditions \cite{McCann}. The application of CNTs, designed to act as the channel of a graphene transistor, will be further discussed in section VI. 

%Using the effective mass model, they  found that both systems  associated with 

\begin{figure}
\begin{center}
\includegraphics[width=8cm]{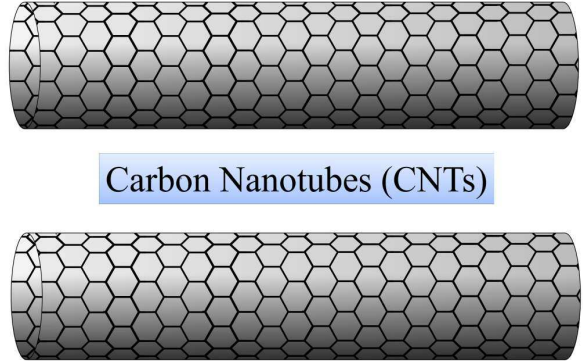}% Here is how to import EPS art
\caption{\label{CNT} The carbon nanotube (CNT) is generally regarded as one-dimensional structure, made by the rolling of graphene into a cylindrical configuration. Its electrical properties (either semiconductor or metallic) are entirely dependent upon the choice of boundary conditions (determined by the chiral index). If the CNT is a semiconductor, the band-gap $E_g$ of graphene is inversely proportional to the nanotube radius $r_{\rm CNT}$.}
\end{center}
\end{figure}			

\begin{figure}
\begin{center}
\includegraphics[width=8cm]{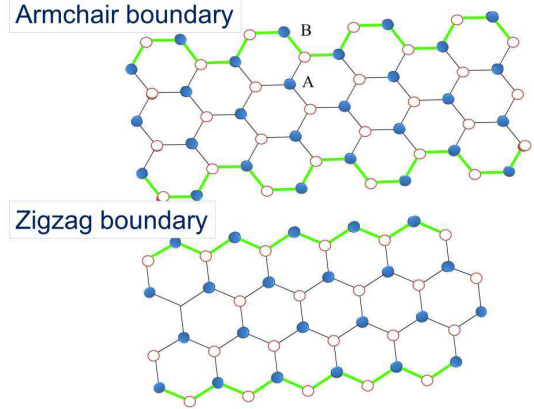}% Here is how to import EPS art
\caption{\label{boundary} The boundary conditions (BCs) applied to the graphene nanoribbon (GNR) will normally determine its conductivity. Both the armchair and zigzag BCs are shown above. When applied to graphene, the armchair BC can infer either a conducting or semiconducting behaviour, whereas the zigzag BC can only be of a conducting nature.}
\end{center}
\end{figure}	

\subsection{Carbon Nanotube}
A carbon nanotube (CNT) can be formed by rolling or twisting a graphene ribbon into a cylindrical configuration \cite{Avouris}. CNTs are often referred to as a one-dimensional nanostructures, due to their tiny radius \cite{Avouris} (cf. Figure~\ref{CNT}). These CNTs can be either single-walled or multi-walled, dependent upon how many monolayers exist \cite{Martel} (for details concerning the synthesis of CNTs, one can refer to \cite{Zhu}). Nevertheless, this folding of graphene leads to deformation of the carbon atoms, thus altering its electronic band properties \cite{Avouris,Tans,Martel}. In general, the band structure of CNTs is determined by the boundary conditions (BCs), specified by the folding angle and radius of the CNT (sometimes known as the chiral index) \cite{Avouris,McCann}. This, in turn, determines whether the CNT is a conductor or a semiconductor. Generally, the boundary of a CNT can be classified as having zigzag or armchair structure, or a mixture of both (cf. Figure~\ref{boundary} for the GNR). Zigzag CNTs are conductors, whereas armchair CNTs can either be a conductor or semiconductor (dependent upon the radius $r_{\rm CNT}$ of the nanotube) \cite{Avouris,Tans,Martel}. This radius is found to be inversely proportional to energy band-gap $E_g$
\begin{equation}
E_g \sim 1/r_{\rm CNT}.
\end{equation} 
The wave function $\psi$ of the CNT is also subject to a periodic boundary condition, thus determining the degeneracy of energy states \cite{Iyechika,Avouris,Tans,Martel}.   
The nanotube can also act as a good transistor channel between the source and drain, since the carriers exhibit long range (ballistic) transport at room temperature \cite{Avouris,Tans,Martel}. Transistor operation and contact effects will be further elaborated upon in section VI.D. 

Single-walled CNTs are a single layer of graphene curled into a hollow cylinder. Recent research by Martel {\it et al.} \cite{Martel}, states that the drain current decreases as the applied gate voltage $V_G$ is increased above the CNT channel. They report that carriers in CNTs come from holes (p-channel) which may either be created due to deformation, or the contacts between the CNT and electrodes \cite{Martel}. For zero gate voltage ($V_G = 0$), a current will flow through the channel. Conversely, for a large positive gate voltage ($V_G > 0$), no current is observed \cite{Martel}. 

Multi-walled CNTs, on the other hand, constitute a roll of multiple layers of graphene. Martel {\it et al.} \cite{Martel} report that no field effect was found, and that multi-walled CNTs are conductive. It is because the radius of a multi-walled CNTs is larger than those of single-walled, that the band-gap is very small \cite{Martel}. 

\subsection{Graphene Nanoribbon}	
A graphene nanoribbon (GNR) is a graphene sheet possessing a very small width $W_{r}$ \cite{Iyechika,Avouris}. A GNR can be obtained by cutting a CNT along its axis, with a scale ranging from 10-100nm \cite{Avouris,Li2}. As with the CNT, the BCs of a GNR determine whether system behaves as a conductor or semiconductor. The zigzag BC corresponds to a conducting GNR, whereas the armchair BC can either yield a conducting or a semiconducting characteristics \cite{Iyechika,Avouris,Li2}. Electrons can become trapped in a small confined region of the GNR with a certain BC, thus forming an energy band-gap \cite{Li2}. For the armchair structure, this band-gap should be (in theory) inversely proportional to the width $W_r$ of the nanoribbon \cite{Iyechika}
 \begin{equation}
 E_g \propto 1/W_r.
 \end{equation} 
Current experiments, however, have shown that the band-gap observed does not rely upon the width and chiral index \cite{Iyechika,Avouris,Tans,Martel,Li2}. It may be plausible that the lithography and etching techniques are inadequate to define the orientation precisely \cite{Avouris,Li2}. Therefore, hybrid structures (combinations of armchair and zigzag GNRs) always co-exist within a sample. In any case, phonon scattering near the edge of the nanoribbon would also downgrade the mobility \cite{Avouris}.  

\begin{figure}
\begin{center}
\includegraphics[width=8cm]{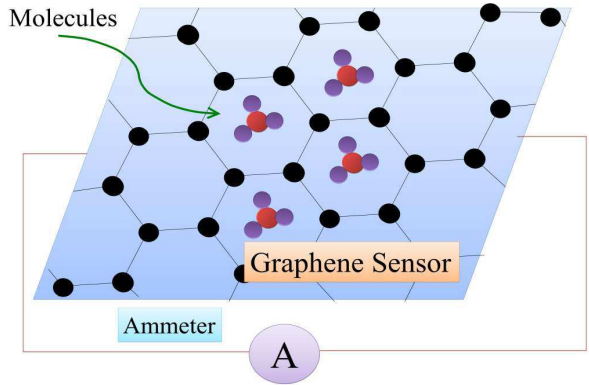}% Here is how to import EPS art
\caption{\label{sensor} A schematic representation of how a graphene sensor may be used as a molecular detector. Graphene is very sensitive to the molecules attached to it, easily affecting its conductivity. In such a case, graphene can be used to detect small biological molecules (such as DNA), or even integrated into fire-alarms as a smoke detection system.}
\end{center}
\end{figure}	

\section{Applications of Graphene Devices}
Some common applications of graphene to electronic devices are discussed below. These include graphene sensors \cite{He,He2}, photonic and optical devices \cite{Iyechika,Engel} and capacitors.  Details of graphene transistors will be discussed in section VI.

\subsection{Graphene Sensor}
Impermeability is one of the most outstanding properties of graphene, since the carbon atoms are closely packed with one another \cite{Savage}. Furthermore, the conductivity of graphene is sensitive to impurities, and can change easily when other molecules or substrates become attached to it. The variation of such conductivity due to molecular attachment is the principal foundation for molecular sensors (cf. Figure~\ref{sensor}) \cite{He2}. Graphene, with its two-dimensional surface, provides a platform upon which tiny particles can attach. The deviation of electrical resistivity, caused by the attached molecules (impurities), can be measured by the Hall Effect \cite{He,He2}. He {\it et al.} \cite{He,He2} have successfully developed bio-sensors and gas sensors using such a method. In the near future, graphene is also likely to be a plausible candidate for magnetoresistance sensors \cite{Tom}. 
	
\subsection{Graphene Photonic Devices}	
There is no doubt that graphene film is an ideal material for photonic devices \cite{Rafiee}. Due to the fact that graphene's atomic thickness has little absorption of light, this makes it almost transparent to all visible wavelengths \cite{O'Hare,Avouris} (cf. section III.C). As such, there are a number of applications, including solar photonic devices, touch screens and graphene photonic transistors \cite{Engel}. 

Optical semiconducting devices are derived from the working principle of the photo-electric effect. Photons transfer their energy to electrons, which can only absorb photons of specific wavelength. Donors are generally doped so as to increase the Fermi level. The absorption range will therefore be reduced, thus enhancing the sensitivity to photon absorption or emission \cite{Engel,Rudden}. Under the influence of an electric field, the emission of photons (due to recombination of the carriers) is referred to as electro-luminescence \cite{Avouris}. Note that some recombination may also relate to the process of thermionic emission. Additionally, photo-conductivity is the process of photon absorption, whereby the conductivity is increased due to the excitation of additional carriers. This process leads to the creation of additional electrons and holes in the conduction and valence bands \cite{Avouris,Rudden}. As such, one could potentially apply this premise to that of light detectors. Indeed, Kim {\it et al.} \cite{Kim2} further noticed that a graphene light detector can have a transit bandwidth reaching speeds of up to $1$THz. 	

\begin{figure}
\begin{center}
\includegraphics[width=8cm]{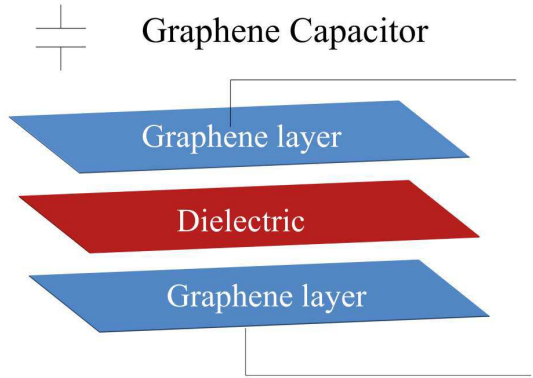}% Here is how to import EPS art
\caption{\label{cap} Here, the graphene layers act as electrodes. This forms the basis of a capacitor which has a fast response time and large capacitance. Furthermore, the size of the capacitor can be minimised due to graphene's atomic thickness. Whilst it is desirable to have a large capacitance $C$, the internal resistance $R$ inside the capacitor is very small, due to the high conductivity of graphene itself. Thus we maintain a small time constant $\tau=RC$ as low as $200\ \mu$s \cite{Miller}. Furthermore, the cost of manufacturing the graphene capacitor is relatively inexpensive.}
\end{center}
\end{figure}	
	
\subsection{Bilayer Graphene Capacitor}		
The working principle of the graphene bilayer capacitor is to store charge and bypass high-frequency signals \cite{Miller}. A simple graphene capacitor consists of two graphene layers, separated by a dielectric medium (cf. Figure~\ref{cap} for a schematic representation). Graphene is ideal for storing charge due to its high mobility. Furthermore, a single atomic layer is also capable of minimising the overall size of the capacitor \cite{Miller}. Interestingly, a graphene capacitor has ambipolar characteristics within the electrodes (the electrodes being either positive or negative), thus simplifying the manufacturing process \cite{Miller}. Miller {\it et al.} \cite{Miller} have also reported that the response time for a capacitor should be fast enough to react to high-frequency circuits. Therefore, the time constant ($\tau=RC$) is made as small as possible, where $R$ is the internal resistance and $C$ is capacitance. It is important to also note that $\tau$ is inversely proportional to cutoff frequency $f_{cut}$. If one requires large capacitance $C$ for charge storage, the internal resistance $R$ inside the capacitor is required to be very small. Graphene, with its low resistance, is therefore an ideal material for the next generation of capacitors. Miller {\it et al.} \cite{Miller} further go on to show that graphene bilayer capacitors can perform extremely well at high frequencies. Thus, the aforementioned graphene devices may be very useful for `green' energy storage. 

\begin{figure}
\begin{center}
\includegraphics[width=8cm]{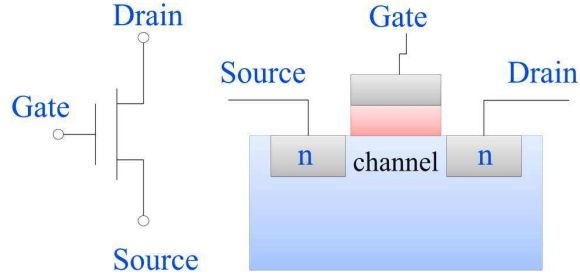}% Here is how to import EPS art
\caption{\label{gate} Schematic of a field-effect transistor (FET) is shown. An applied gate voltage $V_{G}$ can control the accumulation of charge carriers inside the transistor channel. Hence, the flow of current from drain to source can be adjusted accordingly. The concept is analogous to that of a conventional tap - the flow of water representing current.}
\end{center}
\end{figure}	

\section{Developments and Hurdles of Graphene Transistors}
We have already highlighted upon some of the major applications of graphene devices in previous section. However, they have also created a firm base for the next generation of the graphene electronics -- graphene transistors.
Transistors (which are made of semiconductors) are a key constituent of many electronic devices. These include inverting amplifiers, or logical controllers in integrated circuits. Generally speaking, a field-effect transistor (FET) consists of three terminals - gate, source and drain (cf. Figure~\ref{gate}). An electric current flows between the source and drain via a channel, and the current flow rate can be controlled via the gate voltage $V_{G}$ \cite{Irwin}. The gate oxide layer is a dielectric between the gate and channel. Impurities are also introduced via doping, to increase the concentration of charge-carriers on the channel \cite{Rudden,Irwin,Turton}. Two types of carriers: p-type (acceptor) and n-type (donor) are commonly used in industry. The materials constituting an FET (source-channel-drain) are typically of either n-p-n type or p-n-p type \cite{Rudden,Irwin,Turton}. 
The scale and speed of silicon-based transistors has undergone leaps in performance over recent decades. Nevertheless, the limit of transistor size has reached an optimal level, with new materials urgently being required to replace silicon. Graphene, a revolutionary new material, looks set to fill this void due to its outstanding electronic properties \cite{Wallace,Avouris}. 
	
\begin{figure}
\begin{center}
\includegraphics[width=8cm]{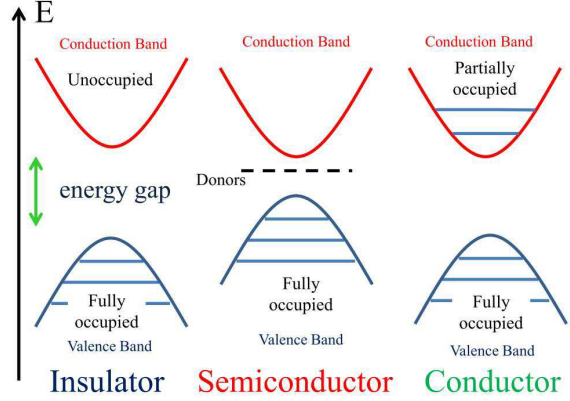}% Here is how to import EPS art
\caption{\label{band} The energy bands for an insulator, semiconductor and conductor are all given. For the insulator, an energy band-gap $E_{g}$ exists. The valence band is fully occupied by electrons, with no occupation of the conduction band. Here, electrons are incapable of moving freely between the bands - as per the definition of an insulator. In the case of a conductor, electrons can move freely within the conduction band (which is partially occupied), and hence current can flow. In particular, a small band-gap exists in the case of a semiconductor. Thus, only a small energy is required to excite an electron from the valance to conduction band. Semiconductors (such as silicon) are generally doped via donors or acceptors in order to make an electron/hole excite more easily across the gap. Using an applied electric field, one can also control the ON/OFF state of a semiconductor using this method.}
\end{center}
\end{figure}	
	
\subsection{Band-gap of Graphene}			
\subsubsection{Insulator, Semiconductor and Conductor}
The electrical conductivity of a material can be split into three categories: insulators, semiconductors or conductors, according to its band structure (cf. Figure~\ref{band}). Electrons are not able to move freely within an insulator, as the valence band (fully occupied by electron states) and conduction band (unoccupied) are separated by an energy band-gap $E_{g}$. The semiconductor, however, is the basic building block for an ON/OFF switching device, as it possesses only a small energy band-gap \cite{Rudden,Irwin,Turton}. For example, silicon is a commonly used material within semiconductors, and is generally doped (via some donor for example) in order to make an electron excite more easily to the conduction band. In semiconductors, electrons can also be activated into the conduction band via the increase of temperature, or an applied electric field (up shift of Fermi level). Conductors normally relate to metals, as electrons are permitted to move freely in the conduction band. It is because metals have this partially filled conduction band, that electron excitation is gapless. 

\subsubsection{Methods of Band-gap Creation in Graphene}	
\begin{figure}
\begin{center}
\includegraphics[width=10cm]{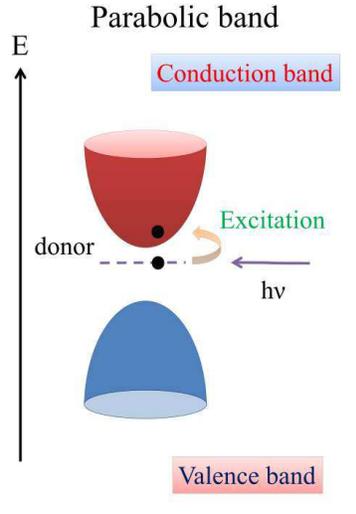}% Here is how to import EPS art
\caption{\label{semiband} The band structure of a semiconductor is shown. Contrary to the linear energy-momentum relationship for graphene, this possesses parabolic shape. A small energy band-gap $E_{g}$ exists, which is crucial for defining the ON/OFF logical state. The diagram also shows how an energy $h\nu$ is required to excite an electron from the valence to conduction band, thus allowing a current to flow.}
\end{center}
\end{figure}

Figure~\ref{semiband} shows that the electronic band structure of a semiconductor is parabolic, with a non-zero energy gap separating the conduction and valence bands \cite{Iyechika,Schwierz}. It is interesting to note that graphene has a linear cone band structure, and is gapless at the Dirac point, (cf. Figure \ref{cone} and section III.D) \cite{Novoselov1,Geim3}. Small thermal fluctuations, or the application of an electric field may excite electrons from the valence to conduction band within graphene \cite{Novoselov1}. 
The Fermi energy will rise in the presence of an applied electric field (dependent upon the sign of the applied gate voltage), and thus exhibit a metallic characteristic. Additionally, there exists an electrical potential of the graphene sheet relative to some carrier reservoir -- this is determined by the geometry of the device.
The zero band-gap of graphene is clearly an obstacle with regards to its application in semiconducting devices, as no precise OFF state is provided \cite{Avouris,Kim2}. Accordingly, opening the band-gap is the most important task in making the graphene transistor become a practical reality. 

Several existing methods to create the gap include carbon nanotubes, graphene nanoribbons, deformed structures and bilayer graphene, to act as a transistor channel \cite{Geim3,Schwierz,Avouris,Kim} (cf. section IV). Graphene oxide is also capable of providing a small energy band-gap due to the change of electronic band structure \cite{Iyechika,He,He2}. Another possibility is via deformation of the graphene layer, either by bending or physical strain, since theoretically, the band-gap properties depend upon the width and orientation of the graphene allotrope \cite{Geim3,Iyechika,Li2}. 
A small band-gap in graphene can also be created via chemical doping, or by retaining some of the defects in the structure as mentioned by Coletti {\it et al.} \cite{Coletti} and Terrones {\it et al.} \cite{Terrones}. As previously discussed, doping in a silicon semiconductor is achieved by replacing some atoms of silicon with other elements. However, this is slightly different for the case of graphene. Due to the strong honeycomb structure, the dopants are generally placed upon the graphene surface, rather than by replacing a carbon atom itself. This leads to a structural defect, and thus changes the electronic band structure. This means that the band-gap can therefore be controlled. Some common dopants include gold, sulphur, boron and nitrogen \cite{Coletti,Terrones}. Boron nitride (BN) is one of the key compounds for the doping of graphene, and subsequent band-gap creation \cite{Fan,Shinde}. Fan {\it et al.} \cite{Fan} have also discovered that BN, doped in a `random' pattern upon the graphene surface, is capable of opening a band-gap. They report that the distribution of surface charges (electrons in $\pi$ bond configuration) is altered by doping, hence forming a small band-gap. However, existing doping methods are sometimes not simple to control \cite{KYan}. However, Yan {\it et al.} \cite{KYan} have recently reported that doping via CVD methods can offer a more simple and stable means of creating this band-gap.  

As discussed in section IV.A, the CNT is a traditional method of opening the band-gap, and is dependent upon the choice of boundary conditions (characterised by the chiral index) \cite{Avouris,Tans,Martel}. Bilayer graphene has a zero band-gap, although an applied electric field can be implemented to create such a gap \cite{Iyechika,Schwierz} (details of the band-gap in bilayer graphene can be referred to in \cite{ Kruczynski,Gradinar}). Deformation of graphene structures has also been known to control the band-gap\cite{Iyechika}. Nevertheless, in all cases, developing a band-gap within graphene would normally reduce its mobility \cite{Iyechika,Schwierz}. 

Schwierz \cite{Schwierz} has stated that the mobility of GNRs only falls within the range of $200-1500$\ cm$^{2}$V$^{-1}$s$^{-1}$ \cite{Schwierz}. The decrease is due to dislocations present in the graphene crystalline structure, most likely induced during the manufacturing process. It is widely known that the speed of the carriers is proportional to their mobility - thus the tradeoff in creating a band-gap is to slow down the speed of the device \cite{Rudden,Turton}. As mentioned by Schwierz \cite{Schwierz}, an ON/OFF switching state is not well-defined in current technologies, and much effort has been focussed upon this area. Alternatively, graphene may be suitable for high-frequency devices, which do not require a well-defined OFF state.

\subsubsection{Band-gap on Substrate in Electric Field}	
As previously mentioned, the most general shape of suspended graphene consists of two sub-lattices (A and B), which can be displaced with respect to each other in a transverse direction \cite{O'Hare} (cf. Figure \ref{Fig1ab}b). If graphene is coated on a substrate, this will give rise to some force (including the van der Waals) upon the interface. The sub-lattices should differ with up and down transverse displacements $u_z$ along with the z-axis, assuming that an electric field $E_z$ is applied. This means that symmetry between the up and down (out-of-plane) displacements is broken. For a single valley, we may estimate the value for the band-gap directly via the 2$\times$2 Dirac Hamiltonian matrix:
 \begin{equation}
\hat{H} = v_{F} {\hat \sigma} \cdot {\bf p} +  \sigma_z\Delta(E_z),\ \ \ \ \ \ \ \ \ \ \sigma_z =\begin{pmatrix}
1 & & 0 \\
0 & & -1
 \end{pmatrix}
\end{equation}
where $ \Delta(E_z) $ describes a stress induced by the substrate, $E_z$ is the electric field in z direction, and $ \sigma_z$ is one of the Pauli matrices. The eigenvalues of this matrix specify the Dirac spectrum $ \epsilon(p)= \pm \sqrt{v_F^2 p^2 %\hbar^2 k^2
+\Delta(E_z)^2}$, with a value for the band-gap equal to $2 \Delta(E_z)$ \cite{O'Hare}.  
 
The application of a transverse electrical field (of amplitude $E_z$) can also induce polarisation of  sublattices (A and B). This is described by the same invariant term in the Hamiltonian as $\sigma_z \Delta(E_z)$. We have $\Delta(E_z)=e E_z u_z$, and thus a band-gap equal to $ 2e u_z E_z$ \cite{O'Hare}. To have visible band-gap, the value of $E_z$ is required to be large. The best way to realise this mechanism, is to use a ferroelectric material for the substrate. However, the deformation phenomena are of utmost importance here \cite{O'Hare}. Due to the transverse lattice displacements separating the sub-lattices (A and B), there arises a deformational field with a value of the order $ D \delta u_z/a$. Here, $D$ is the deformation potential, `$a$' is the inter-atomic spacing, and $\delta u_z$ is the change of transverse displacement which arises due to the presence of the substrate. In the following estimation for the energy band-gap, we use the following parameters for graphene, $a=1.4\AA$ and the deformation potential is equal to $D=6.8$ eV \cite{Kaasbjerg}. If we assume that $\delta u_z$ is around 10\% of the transverse displacement $u_z$, which is roughly equal to $0.4\AA$ (the order of magnitude estimated for silicene \cite{Drummond}), this gives an energy band-gap $E_g \approx 190$ meV \cite{O'Hare,Drummond}.
Thus, in the addition to deformation from the substrate, the application of $E_z$ can contribute an additional force to the substrate, and may provide a further means of controling the band-gap. This value can increase or decrease linearly with the applied electric field. The phenomenon is general, and also arises in other two-dimensional crystals (cf. for example, silicene \cite{O'Hare,Drummond}). 

\subsection{Electric Field Effect in Graphene}		
\begin{figure}
\begin{center}
\includegraphics[width=8cm]{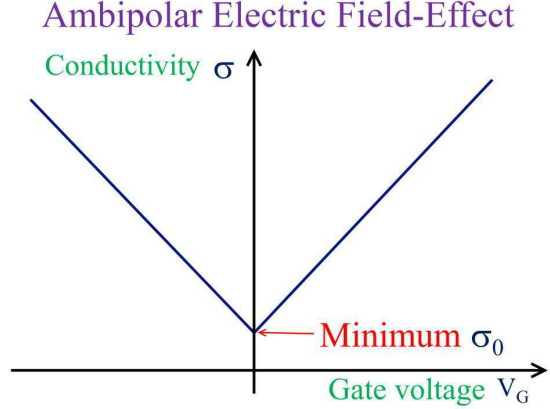}% Here is how to import EPS art
\caption{\label{bipolar} The ambipolar characteristics of the electric field effect in graphene were first observed by Geim {\it et al.} \cite{Geim1}. It is one of the key important properties that the conductivity $\sigma$ increases with the absolute value of $V_G$ (associated with the DoS). There is also a minimum quantum conductivity $\sigma_0$, which exists in the absence of an applied gate voltage (i.e., $V_{G}=0$).}
\end{center}
\end{figure}

Geim {\it et al.} \cite{Novoselov1} have observed that electric field effects were found in two-dimensional graphene, and that such effects were stable at room temperature. Figure~\ref{bipolar} shows the schematic behavior of electrical conductivity in the presence of a gate voltage $V_G$. A minimum conductivity $\sigma_{0}$ is also shown to exist at the Dirac point. Although there exists a zero charge density at this point, nevertheless, it exhibits metallic properties \cite{Novoselov1}. A linear relationship between conductivity and gate voltage is found for an uncontaminated graphene film, with positive and negative slopes when $V_G > 0$ and $V_G < 0$ respectively \cite{Geim3,Novoselov2}. This indicates that conductivity is proportional to the DoS at the Fermi energy. An ambipolar characteristic within the electric field has also been discovered, and that either holes ($V_G < 0$) or electrons ($V_G > 0$) could both act as transport carriers \cite{Fal'ko}. 

In the early studies of Geim {\it et al.} \cite{Novoselov1}, hole carriers were found to exist even at $V_G = 0$. Here, water vapor molecules attach to the surface of the film, thus changing the polarity. These water molecules can, however, be removed via the process of annealing \cite{Novoselov2}. Recent reports \cite{Geim1,Geim2,Neto} also reveal that the mobility is stable when subject to a change of temperature. Only structural defects can affect the mobility of a sample - this would not reduce significantly, even in the presence of high electric fields. A monolayer graphene channel would also appear to remedy the short-channel effect present in silicon-based transistors. Such an improvement is one of the key foundations for a faster logical transistor. 

Schwierz \cite{Schwierz} also points out that a graphene transistor might have some special $I-V$ characteristics, possessing two linear regions and one saturation region. When the drain-source voltage $V_{DS}$ is small, a linear relationship for the $I-V$ curve is found. As $V_{DS}$ is increased further, a saturation of the $I-V$ curve begins to emerge. If $V_{DS}$ exceeds a certain limit, a new linear region is yet again observed. Only a few studies with current saturation have been reported. A greater understanding of $I-V$ characteristics would require addressing in the near future, as this saturation region would affect the performance of devices. 
 
\subsection{Graphene Oxide Transistor}	
\begin{figure}
\begin{center}
\includegraphics[width=8cm]{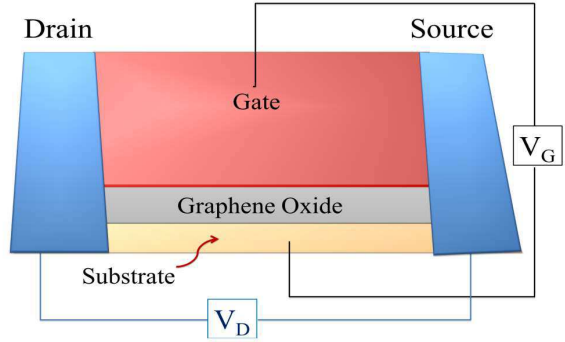}% Here is how to import EPS art
\caption{\label{tranoxide} Reduced graphene oxide (rGO) is a compound of both graphene and oxygen, possessing a very small band-gap. This acts as the channel inside the transistor. Despite the mobility of graphene oxide being less than that for pure graphene, it is one of the simplest ways to manufacture graphene transistors in industry.}
\end{center}
\end{figure}

A number of conventional graphene transistors, of various design, have been proposed in recent years. In principle, graphene can easily act as a transistor channel \cite{Avouris}. Graphene oxide (GO) is a compound of both graphene and oxygen \cite{Geim2,Iyechika}, a single layer of which is slightly thicker than a graphene monolayer \cite{Jin}. Jin {\it et al.} \cite{Jin} claim that the use of graphene oxide is one of the simplest ways to create an energy band-gap, thus turning graphene into a semiconductor. Graphene oxide is water soluble, and the fabrication process is not at all complicated. Recent studies \cite{Jin} have reported that graphene oxide can be tuned to either a semiconducting or insulating state, dependent upon the amount of oxygen, and bonding structure therein. A thin graphene oxide layer is generally coated on a silicon substrate, acting as a channel for the transistor. A schematic of a graphene oxide transistor is shown in Figure~\ref{tranoxide}, and exhibits an ambipolar characteristic for a clean sample \cite{Jin}. It was also discovered that the drain-source current increases with the absolute value of gate voltage \cite{Jin}. The drain-source current was also found to be increasing with temperature. Despite the mobility of graphene oxide being much less than for graphene ($\sim 100\ $cm$^{2}$V$^{-1}$s$^{-1}$), the main advantage of this method is its simplicity, requiring no reduction process (cf. Jin {\it et al.} \cite{Jin}). 
	
Reduced graphene oxide (rGO) is obtained via a chemical process of removing oxygen from graphene oxide \cite{Iyechika}. However, reduction is not always perfect. Defects in the honeycomb structure are usually found in experiments \cite{Iyechika,Dreyer,Jin}, with residual oxygen compounds sometimes remaining on the boundary of samples \cite{He,He2,Jin}. High temperatures can remove the oxide components, but are required to be in excess of 1000K. Like graphene oxide, reduced graphene oxide can act as the channel inside an FET \cite{He2}. He {\it et al.} \cite{He2} have applied a coating of reduced graphene oxide to polymer substrates (3-aminopropyltriethoxysilane PET substrate), discovering an ON/OFF switching ratio of approximately 3-4. This switching ratio is determined by how the current responds to the gate voltage - from the OFF to ON state, or vice versa. They further reveal that the conductivity of a transistor depends upon the thickness of the rGO electrodes - the mobility also degrading as a result of structural defects. However, even in the absence of a gate voltage, a non-zero drain-source current still exists, and thus current leakage may occur. He {\it et al.} \cite{He2} also mention that a bending of reduced graphene oxide would widen the transition region of the ON/OFF state, thus reducing the switching ratio. 
 
\begin{figure}
\begin{center}
\includegraphics[width=8cm]{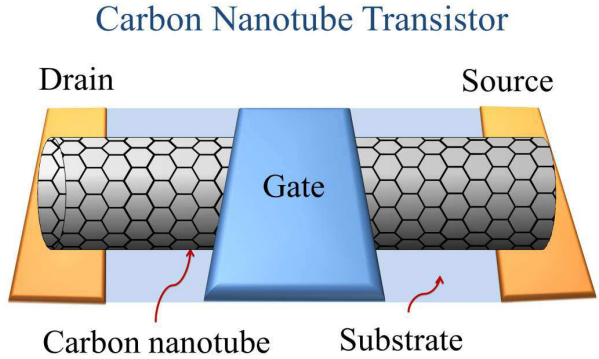}% Here is how to import EPS art
\caption{\label{tranCNT} The CNT transistor is a well-developed area of research. A band-gap can be created by altering the boundary conditions. Normally, the performance of a CNT transistor will be characterised not only by the resistance $R$, capacitance $C$, and inductance $L$ of the channel, but also the contact interfaces of the metal electrodes.}
\end{center}
\end{figure}	

\subsection{Carbon Nanotube Transistor}	
A number of theories and experiments with CNT transistors \cite{Schwierz,Avouris,Tans,Kreupl,Franklin,Jang} have been undertaken for a few decades now, becoming a well-developed area of research. The CNT works as a transistor channel, capable of being tuned from conductor to semiconductor as mentioned by Tans {\it et al.} \cite{Tans}. Figure~\ref{tranCNT} shows a diagrammatic representation of such a CNT transistor. In theory, the CNT channel can perform long-range (ballistic) electrical transport \cite{Schwierz}. Avouris {\it et al.} \cite{Avouris} have also proposed that the movement of electrons within the CNT structure is restricted to one dimension \cite{Avouris}, thus limiting the amount of inelastic scattering. In brief, the performance of a CNT transistor can be characterised by its resistance $R$, capacitance $C$ and inductance $L$ \cite{Avouris}. In the studies of Avouris {\it et al.} \cite{Avouris}, they further mention that these physical quantities ($L$, $R$ and $C$) simultaneously exhibit both classical and quantum behavior - these quantum effects are found to be more important for a very small transistor size. In fact, the quantum resistance $R_q$ is of the scale $h/4e^2$, whilst quantum capacitance $C_q$ and inductance $L_q$ in CNTs are associated with the DoS (cf. for details \cite{Schwierz,Avouris}). Capacitance and inductance, whether classical or quantum, depend upon the shape, size and boundary conditions of the CNT. Moreover, Schottky barriers are also found in CNT transistors, exhibiting a large resistance. These are due to the contact between the metal electrode and semiconducting CNT. Due to the differing work functions between the CNT and metal electrode, this leads to a consequent shift of the Fermi energy \cite{Avouris}. Recently, it has been discovered that the shape of potential barrier, between the electrodes and CNT, is shallow and wide. They maintain quantum coherence of electron transport between the adjacent sections, for lengths even up to several micrometers \cite{Barbara}.
	
Theoretically, ballistic transport is found throughout the CNT transistor, with a large mean-free-path of the carriers \cite{Schwierz,Avouris}. In the real case, like in metals and semiconductors, the electron transport is due to diffusion, thermal gradients and electric fields, together with the chiral index of CNT. The length of the CNT also determines the transport - the longer the nanotube, the more inelastic scattering that will occur \cite{Schwierz,Avouris}. The drain-source current can be controlled via an electric field acting upon the CNT channel. Two types of CNT transistor (top gate and wrap-around gate) were developed recently \cite{Schwierz,Avouris}. Avouris {\it et al.} \cite{Avouris} have pointed out that the Schottky barriers in CNT-metal electrode contacts are very high when compared to those of silicon. Nonetheless, CNT transistors can work well upon inter-connection of intergated circuits, operating well even at room temperature \cite{Avouris}. This is due to the small electron-phonon interaction inside the CNT. Overall, these transistors consume less power than their traditional silicon-based counterparts, thus proving to be an advantage for logic switches.
	
Franklin {\it et al.} \cite{Franklin} have simulated a short CNT transistor (with 9nm channel width), with reports of better performance and stable electrical properties in comparison to silicon-based devices possessing a 10nm channel width. The study claims that the short CNT channel provides a faster switching speed. However, in a silicon-based transistor, a shorter channel would deteriorate its quality. The overall stability of the drain-source current is compromised when the channel is of the scale of tens of nanometers (i.e., similar to that of the depletion layer). The electrical current through the transistor thus becomes unreliable \cite{Avouris}. This is more commonly known as the short-channel effect. According to Franklin {\it et al.} \cite{Franklin}, the short-channel effect can be minimised by a thin CNT. In addition, Jang {\it et al.} \cite{Jang} have also reported that CNTs provide a high switching ratio. 

\begin{figure}
\begin{center}
\includegraphics[width=8cm]{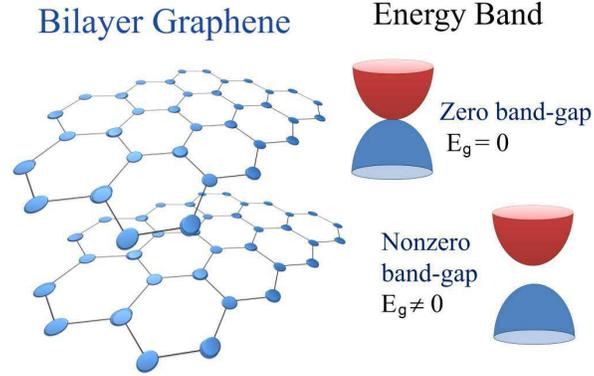}% Here is how to import EPS art
\caption{\label{bilayer} The band structure of bilayer graphene has a parabolic shape, similar to that of a semiconductor. However, bilayer graphene is gapless, whereas the band-gap of a semiconductor is non-zero. In the presence of an electric field, a band-gap in graphene can be created and adjusted accordingly. More importantly, its bilayeral structure can reduce current leakage.}
\end{center}
\end{figure}

\subsection{Bilayer Graphene Transistor}	
Bilayer graphene consists of two individual graphene monolayers (cf. Figure~\ref{bilayer}). As expected, Sarma {\it et al.} \cite{Sarma} have pointed out that the structure of bilayer graphene is similar to that of very thin graphite. Similar to monolayer graphene, the bilayer has a zero energy band-gap, but with a parabolic shape \cite{Iyechika}. A perpendicular electric field applied to the graphene bilayer, or impurities via doping can, however, create a band-gap \cite{Iyechika}. Furthermore, the magnitude of this electric field can tune the size of the gap \cite{Zhang3}. Ponomarenko {\it et al.} \cite{Ponomarenko}  have also reported that the bilayer channel can be changed from conductor to insulator by tuning the applied electric field. However, one cannot control the band-gap easily via doping \cite{Iyechika}. The mobility in double layers is lower than that of a single layer, as stated by Wallace \cite{Wallace}. Nonetheless, carriers in bilayer channels can still perform at high speed. Studies by Zhang {\it et al.} \cite{Zhang3} have discovered that the band-gap of bilayer graphene varies in the range of 0--298\,meV at room temperature. This narrow gap is suitable for tunable nanophotonic applications, allowing a certain range of the visible spectrum to pass through \cite{Zhang3}. Another advantage of bilayer graphene, is that the current leakage can be minimal. A more concise understanding as to precisely how the band-gap of bilayer or multilayer graphene is affected by an external electric field, will urgently be required in the near future.   

\begin{figure}
\begin{center}
\includegraphics[width=8cm]{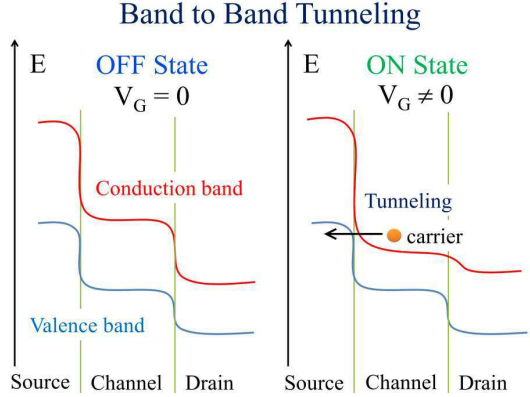}% Here is how to import EPS art
\caption{\label{trantunnel} The idea of the tunneling transistor is shown. Tunneling occurs during the transition of an electron from the conduction to valence band, as a gate voltage $V_G$ is applied (the idea is similar to that of Landau-Zener tunneling). The gate voltage could alter the shape of the conduction band, and thus control the tunneling current between the two bands. A tunneling transistor generally consumes much less power than its traditional silicon-based counterpart.}
\end{center}
\end{figure}

\subsection{Tunneling Graphene Transistor}	
Figure~\ref{trantunnel} shows the ON/OFF states of a transistor, with and without an applied gate voltage $V_{G}$. As mentioned by Ionescu {\it et al.} \cite{Ionescu}, the application of a gate voltage could control tunneling between the two bands. A tunneling graphene transistor possesses a steep sub-threshold slope, hence providing a fast response time \cite{Schwierz}. The basic principle is that the applied gate voltage alters the shape (offset) of the conduction band, and hence the charge carriers from the conduction band could tunnel to the valence band. The choice of material for the channel inside is usually composed of either CNTs, GNRs or bilayer transistors \cite{Dean}. Ionescu {\it et al.} \cite{Ionescu} state that the current in tunneling transistors arises due to a leap of electrons between the conduction and valence bands, rather than via diffusive and thermal transport (the concept is similar to Landau-Zener tunneling \cite{Nandkishore}). In general, a tunneling transistor (via some quantum process) consumes much less power than a non-tunneling equivalent \cite{Ionescu}. Michetti {\it et al.} \cite{Michetti} have also proposed that within the tunneling graphene transistor, the ON/OFF switching ratio could reach as high as $10^{4}$, and that tunneling could occur even with a small applied gate voltage. Overall, the band-to-band tunneling in a graphene transistor is generally dependent upon the band-gap width, channel length, oxide thickness and gate voltage \cite{Michetti}. 

Zhao {\it et al.} \cite{Zhao} have studied a nanoribbon tunneling transistor, claiming that an ambipolar behavior can be altered to an asymmetric type via doping - highly suited to logic switches. Their study \cite{Zhao} also notes that a subthreshold slope was discovered of 14mV/dec (i.e., high switching ratio), thus consuming less electrical power. 
Another report by Zhang {\it et al.} \cite{Zhang2} has also revealed that the nanoribbon tunneling transistor could also operate at high speed. The tunneling probability is a function of the energy band-gap $P(E_g)$, that is 
\begin{equation}
P(E_g) \propto \exp[{-k{E_g}^2d/(\hbar v_{F} \,e V_{DS})}], 
\end{equation}
where $k$ is a dimensionless constant, $d$ refers to the channel width, $e$ is an electric charge, while $V_{DS}$ is the drain-source voltage \cite{Zhang2}. The width of the ribbon would also affect tunneling probability. Zhang {\it et al.} \cite{Zhang2} further explain that a leakage current is present in the OFF state, most likely associated with some thermal emission. However, Yang {\it et al.} \cite{Yang}, in their study, claim that the nanoribbon tunneling transistor consumes energy which is 8-9 orders of magnitude lower than a silicon-based counterpart.

\subsection{Vertical Tunneling Graphene Transistor}	
Malec {\it et al.} \cite{Malec} have investigated the vertical tunneling between graphene and adjacent metal layers. They found that the DoS in graphene would consequently affect the conductivity of the devices \cite{Malec}. Cobas {\it et al.} \cite{Cobas} have also studied graphene as a dielectric sandwich, positioned between two metal plates. Quantum tunneling was found to act perpendicular between the two plates, and is dependent upon temperature \cite{Cobas}. In 2012, Britnell {\it et al.} \cite{Britnell} proposed a new type of transistor that stems from the idea of quantum tunneling between the insulating and graphene layers, thus effectively acting as a vertical transistor channel. 

The multilayer tunneling transistor \cite{Britnell} constitutes a few insulating layers of hexagonal boron nitride (hBN), positioned between the two graphene plane electrodes. Thus, the hBN is a kind of dielectric barrier which can prevent current leakage, thus consuming less energy.

\begin{figure}
\begin{center}
\includegraphics[width=8cm]{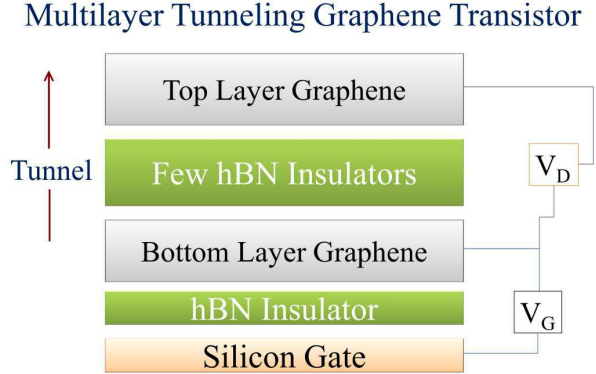}% Here is how to import EPS art
\caption{\label{tranvert} The multilayer graphene tunneling transistor, as proposed by Britnell {\it et al.} \cite{Britnell}. Its innovative design leads to electrons tunnelling vertically, rather than horizontally across the channel. Both the gate and drain-source voltage are applied perpendicular to the layers, with charge carriers subsequently accumulating on graphene layers. Electrons can then tunnel between these layers.}
\end{center}
\end{figure}

The underlying mechanism \cite{Britnell} is as follows (cf. Figure~\ref{tranvert} for a diagrammatic representation). The Fermi levels and potential barriers are controlled by an applied gate voltage $V_G$. An electron tunnels between the two graphene layers, only when an electric field (drain-source voltage) is applied, and is sufficiently high to overcome the barrier. When the gate voltage is applied between the silicon gate and the bottom graphene layer, charge carriers in the top and bottom graphene layers begin to accumulate. The Fermi level is raised, and the DoS increases at the same time. Together with the potential difference between the two graphene layers, electrons will be more likely to tunnel form one graphene layer to another. Not only does the applied gate voltage change the DoS, but also modifies the potential barriers between the graphene and insulating layers \cite{Britnell}. Therefore, a tunneling current through hBN layers is induced, and can be tuned by the magnitude of $V_G$. 
Britnell {\it et al.} \cite{Britnell} have also found that the transport carriers are holes. The report also reveals that a larger $V_G$ and separation of the layers would also enhance the switching ratio. The tunneling nature would also allow the transistor to operate at high frequency, as well as with a high current-gain (cf. for details, Georgiou {\it et al.} \cite{Georgiou}). 

\subsection{Radio Frequency Transistor}	
One can create a band-gap in graphene via any of the aforementioned methods. Even if the band-gap is open, generally, the switching ratio is very small. This is one of the major drawbacks of graphene logic transistors. However, this is no problem for an operational amplifier, or in high-frequency circuits \cite{Schwierz,Kim2}. An amplifying transistor, in general, only requires an ON state. This is because an amplifier requires only a current/voltage input for its operation - thus no OFF state is required. \cite{Schwierz}. The high mobility of carriers in graphene renders a fast response to the input signal, which can be up to a few hundred GHz, (cf. for details \cite{Kim2,Lin,Wu2,Zheng}). More specifically, the cut-off frequency $f_{\rm cut}$ and the maximum oscillation frequency $f_{\rm osc}$ both determine the speed of a transistor \cite{Kim2}. This maximum oscillation frequency refers to the unity power-gain \cite{Wu2}. The cut-off frequency, however, is defined by a current-gain equal to one \cite{Wu2}, and is proportional to the trans-conductance $g_{rf}$.
\begin{equation}
f_{\rm cut} \propto g_{\rm rf}/L_{\rm gate}.
\end{equation}
where $L_{\rm gate}$ is the transistor gate length. To reach a high value for $f_{cut}$, one has to maximise $g_{\rm rf}$ and minimise $L_{\rm gate}$ \cite{Schwierz,Kim2,Wu2}. Wu {\it et al.} \cite{Wu2} have studied the shortest gate length down to 40nm with a cut-off frequency of 155\ GHz. Zheng {\it et al.} \cite{Zheng} have also found, via simulation, that the maximum cut-off frequency can reach up to 300\ GHz. On the other hand, impurities and defects would not only downgrade the value of $f_{cut}$, but the maximum oscillation frequency as well \cite{Kim2}. Wu {\it et al.} \cite{Wu2} also mention that the power-gain in the graphene transistor is particularly low because of the unstable saturation region, of which at present, not much is understood. They add that the maximum oscillation frequency $f_{\rm osc}$ would generally be affected by the size and configuration of the device. Wu {\it et al.} \cite{Wu2} have also reported that $f_{\rm osc}$ can reach upwards of 20\ GHz. It is important to note that the speed of the graphene transistor is influenced by the saturation of the drain current (cf. recent studies by \cite{Schwierz,Wu2,Meric}). As previously mentioned, the contact resistance between the metallic materials and graphene is hugely problematic, and even more serious than the undefined saturation drain current \cite{Kim2, Wu2}. Researchers are looking for further developments to minimise this resistance.  

\section{Further Developments and Conclusion}
Indeed, a long road lies ahead to implement electronic graphene devices into industry. One of the main reasons is that we cannot maintain both the quantity and quality of graphene devices simultaneously throughout the manufacturing processes. For example, mechanical exfoliation can produce high quality graphene flakes, but only a small amount each time. Chemical vapor deposition and evaporation of silicon carbide are able to mass produce graphene substrates - however, these methods require very high temperatures which may sometimes burn out the graphene structure. It is of utmost importance to develop more precise fabrication techniques to reduce the impurities and defects of graphene products \cite{ Novoselov5}. Most current research is based in laboratories at huge financial expense. How to minimise the spending cost is the next issue to be overcome. 

It is highly important to open up a band-gap for the graphene switching devices, of which existing methodologies include carbon nanotubes \cite{Behabtu}, graphene nanoribbons, deformed structures and bilayer graphene. However, this normally leads to a downgrading of the mobility. One of the major problems for the graphene transistor is current leakage, and the absence of a well-defined ON/OFF switching ratio. The tunneling transistor \cite{Britnell} may be an innovative solution to this problem, in that it consumes less power and operates at a faster speed. Graphene can also be used in high-frequency amplifiers which do not require a well-defined switching ratio. A recent theory reveals that speeds could even reach upwards of 1THz for photonic devices \cite{Kim2}, although one must be careful with regards to the huge contact resistance and unknown saturation regions of a graphene transistor. Some new concepts, such as the single electron transistor and quantum dot transistor, are of great interest, and are being applied to graphene in current research \cite{Behabtu,Lansbergen,Westervelt,Ihn}. Some research groups are now looking for more stable two-dimensional materials, such as graphyne and silicene \cite{O'Hare,Malko,Drummond}, both of which also possess very good mobility. Both O'Hare {\it et al.}\cite{O'Hare} and Drummond {\it et al.} \cite{Drummond} further mention that the band-gap in silicene can be tuned by an applied electric field. Nonetheless, for now, graphene remains a the forefront for understanding the nanoscale effects of the future generation of transistors \cite{Britnell2}.


\begin{thebibliography}{0}
%[1]	Electric field effect in atomically thin carbon films 
\bibitem{Novoselov1} K. S. Novoselov {\it et al.}, {\it Electric field effect in atomically thin carbon films}, Science  306 (2004), pp. 666-669.
%[2]	Graphene: status and prospects
\bibitem{Geim1} A.K. Geim, {\it Graphene: status and prospects}, Science 324 (2009), pp. 1530-1534.
%[3]	Graphene prehistory
\bibitem{Geim2} A. K. Geim, {\it Graphene prehistory}, Phys. Scr. T146 (2012), p.014003. 
%[4]	The rise of graphene
\bibitem{Geim3} A. K. Geim and  K. S. Novoselov, {\it The rise of graphene},  Nature Mater.  6 (2007), pp. 183-191. 
%[5]	The structure of suspended graphene sheets
\bibitem{Meyer} J. C. Meyer, {\it The structure of suspended graphene sheets}, Nature 446 (2007), pp. 60-63. 
%[6]A stable flat form of two-dimensional crystals: could graphene, silience, germanene be minigap semiconductors?
\bibitem{O'Hare} A. O'Hare, F. V. Kusmartsev and K. I. Kugel, {\it  A stable flat form of two-dimensional crystals: could 
graphene, silicene, germanene be minigap semiconductors}, Nano Lett. 12 (2012), pp. 1045-1052.
%[7]
\bibitem{O'Hare1} A. O'Hare, F. V. Kusmartsev, K. I. Kugel, {\it Stable forms of two-dimensional crystals and graphene}, Physica B-Cond. Matt., 407 (2012), pp.1964-1968.   
%[8]Breakdown of continuum mechanics for nanometre-wavelength rippling of graphene
\bibitem{Tapaszto} L. Tapaszto, T. Dumitrica, S. J. Kim, P. Nemes-Incze, C. Hwang and L. P. Birol, {\it  A stable flat form of two-dimensional crystals: could graphene, silicence, germanene be minigap semiconductors}, Nature Phys. 8 (2012), pp. 739-742.
%[9] Electrically Tunable Band Gap in Silicene
\bibitem{Drummond} N. D. Drummond, V. Zolyomi and V. I. Fal'ko, {\it  Electrically Tunable Band Gap in Silicene}, Phys. Rev. B 85 (2012), p. 075423.
%[10]	Quantum information on chicken wire
\bibitem{Fal'ko} V. Fal'ko, {\it Graphene: Quantum information on chicken wire}, Nature Physics 3 (2007), pp. 151-152. 

%[11]Two-dimensional gas of massless dirac fermions in graphene
\bibitem{Novoselov2} K. S. Novoselov, {\it Two-dimensional gas of massless Dirac fermions in graphene}, Nature 438 (2005), pp. 197-200. 
%[12]New directions in science and technology: two-dimensional crystals
\bibitem{Neto}  A. H. C. Neto and  K. S. Novoselov, {\it New directions in science and technology: two-dimensional crystals}, Rep. Prog. Phys. 74 (2011), p. 082501 . 
%[13]Thermal conductivity of isotopically modified graphene
\bibitem{Chen} S. Chen, {\it Thermal conductivity of isotopically modified graphene}, Nature Mater. 11 (2012), pp. 203-207. 
%[14] Wetting transparency of graphene
\bibitem{Rafiee} J. Rafiee, {\it Wetting transparency of graphene}, Nature Mater. 11(2012), pp. 217-222 .
%[15] Room-temperature quantum hall effect in graphene
\bibitem{Novoselov3}  K. S. Novoselov {\it et al.}, {\it Room-temperature quantum hall effect in graphene}, Science 315 (2007), p. 1379.
%[16] Graphene: carbon in two dimensions
\bibitem{Katsnelson}  M. I. Katsnelson, {\it Graphene: carbon in two dimensions}, Materials Today 10 (2007), pp. 20-27.
%[17] Relativistic Brownian motion on a graphene chip
\bibitem{Pototsky} A. Pototsky, F. Marchesoni, F.V. Kusmartsev, P. Hanggi and S. E. Savel'ev, {\it Relativistic Brownian motion on a graphene chip}, European Physical Journal B 85 (2012), p. 356.
%[18] Spectrum of localized states in graphene quantum dots and wires
\bibitem{Zalipaev} V. V. Zalipaev, D. N. Maksimov, C. M. Linton and  F.V. Kusmartsev, {\it Spectrum of localized states in graphene quantum dots and wires}, Phys. Lett. A 377 (2013), pp. 216-221.
%[19] Multicomponent fractional quantum hall effect in graphene
\bibitem{Dean} C. R. Dean {\it et al.}, {\it Multicomponent fractional quantum hall effect in graphene}, Nature Phys. 7 (2011), pp. 693-696.
%[20] Spin qubits in graphene quantum dots
\bibitem{Trauzettel} B. Trauzettel, D. Bulaev, D. Loss and G. Burkard, {\it Spin qubits in graphene quantum dots}, Nature Phys.  3 (2007), pp. 192-196.


%[21] Theory and Design of Quantum Coherent Structures
\bibitem{Zagoskin} A. M. Zagoskin, {\it Quantum Engineering: Theory and Design of Quantum Coherent Structures.} Cambridge University Press (2011).
%[22] Application of graphene to high-speed transistors: expectations and challenges
\bibitem{Iyechika} Y. Iyechika, {\it Application of graphene to high-speed transistors: expectations and challenges}, Science and Technology Trends - Quarterly Review  37 (2010), pp. 76-92.
%[23] Super Carbon
\bibitem{Savage} N. Savage, {\it Super Carbon}, Nature 483 (2012), S30-S31.
%[24] Band theory of graphite
\bibitem{Wallace} P. R. Wallace, {\it Band theory of graphite}, Phys. Rev. 71 (1947), pp. 622-634.
%[25] From conception to realization: an historical account of graphene and some perspectives for its future
\bibitem{Dreyer} D. R. Dreyer, R. S. Ruoff and C. W. Bielawski, {\it From conception to realization: an historical account of graphene and some perspectives for its future}, Angew. Chem. Int. Ed. 49 (2010),  pp. 9336-9344.
%[26] Hochstlamellarer Kohlenstoff aus Graphitoxyhydroxyd
\bibitem{Ruess} V. G. Ruess and F. H. Vogt, {\it Hochstlamellarer Kohlenstoff aus Graphitoxyhydroxyd}, Monatshefte fr Chemie 78 (1948), pp. 222-242.
%[27] Das Adsorptionsverhalten Sehr Dunner Kohlenstoff-Folien 
\bibitem{Boehm1} H. P. Boehm, A. Clauss, G. O. Fischer and U. Hofmann, {\it Das Adsorptionsverhalten Sehr Dunner Kohlenstoff-Folien},  Z. Anorg. Allg. Chem. 316 (1962), pp. 119-127.
%[28] Dunnste Kohlenstoff-Folien
\bibitem{Boehm2} H. P. Boehm, A. Clauss, U. Hofmann and G. O. Fischer, {\it Dunnste Kohlenstoff-Folien}, Zeitschrift Fur Naturforschung Part B-Chemie Biochemie Biophysik Biologie Und Verwandten Gebiete. B 17 (1962), pp. 150-153.
%[29]Two dimensional graphite films on metals and their intercalation
\bibitem{Gall} H. R. Gall, E. V. Rutkov and A. Y. Tontegode, {\it Two dimensional graphite films on metals and their intercalation}, Int. J. Mod. Phys. B 11 (1997), pp. 1865-1911.
%[30]Ultrathin Epitaxial Graphite: 2D Electron Gas Properties and a Route toward Graphene-based Nanoelectronics
\bibitem{Berger} C. Berger {\it et al.}, {\it Ultrathin epitaxial graphite: 2d electron gas properties and a route toward graphene-based nano-electronics}, J. Phys. Chem. B 108 (2004), pp. 19912-19916.


%[31]Magneto-spectroscopy of epitaxial few-layer graphene
\bibitem{Sadowski} M. L. Sadowski, G. Martinez, M. Potemski, C. Berger and W. A. de Heer, {\it Magneto-spectroscopy of epitaxial few-layer graphene} Solid State Comm. 143 (2007), pp. 123-125.
%[32]Two-dimensional atomic crystals
\bibitem{Novoselov4} S. Novoselov {\it et al.}, {\it Two-dimensional atomic crystals}, PNAS 102 (2005), pp. 10451-10453.
%[33] The transistor, a semiconductor triode
\bibitem{Bardeen} J. Bardeen and W. H. Brattain, {\it The transistor, a semiconductor triode}, Phys. Rev. 74 (1948), pp. 230-231.
%[34] Graphene transistors
\bibitem{Schwierz} F. Schwierz, {\it Graphene transistors}, Nature Technology 5 (2010), pp. 487-496.
%[35]Towards wafer-size graphene layers by atmospheric pressure graphitization of silicon carbide
\bibitem{Emtsev} K. V. Emtsev {\it et al.}, {\it Towards wafer-size graphene layers by atmospheric pressure graphitization of silicon carbide}, Nature Mater. 8 (2009), pp. 203-207.
%[36]Epitaxial graphene
\bibitem{Heer} W. A. de Heer {\it et al.}, {\it Epitaxial graphene}, Solid  State Comm. 143 (2007), pp. 92-100.
%[37]Three-dimensional flexible and conductive interconnected graphene networks grown by chemical vapor deposition
\bibitem{Chen2} Z. P. Chen, W. Ren, L. Gao, B. Liu, S. Pei and H. Cheng, {\it Three-dimensional flexible and conductive interconnected graphene networks grown by chemical vapor deposition}, Nature Mater. 10 (2011), pp. 424-428.
%[38]Large-area synthesis of high-quality and uniform graphene films on copper foils
\bibitem{Li} X. Li {\it et al.}, {\it Large-area synthesis of high-quality and uniform graphene films on copper foils}, Science 324 (2009), pp. 1312-1314.
%[39]Large-scale pattern growth of graphene films for stretchable transparent electrodes
\bibitem{Kim} K. S. Kim {\it et al.}, {\it Large-scale pattern growth of graphene films for stretchable transparent electrodes}, Nature 457 (2009), pp. 706-710.
%[40]Imaging ultra-thin layers with helium ion microscopy: utilizing the channeling contrast mechanism
\bibitem{Hlawacek} G. Hlawacek {\it et al.}, {\it Imaging ultra-thin layers with helium ion microscopy: utilizing the channeling contrast mechanism}, Beilstein J. Nanotechnol.  3 (2012), pp. 507-512.


%[41]Correlating raman spectral signatures with carrier mobility in epitaxial graphene: a guide to achieving high mobility on the wafer scale
\bibitem{Robinson} J. A. Robinson {\it et al.} {\it Correlating raman spectral signatures with carrier mobility in epitaxial graphene: a guide to achieving high mobility on the wafer scale}, Nano Lett. 9 (2009), pp. 2873-2876.
%[42] Raman Spectroscopy of four epitaxial graphene layers: Macro-island grown on 4H-SiC (000) substrate and an associated strain distribution
\bibitem{Trabelsi} A. Ben Gouider Trabelsi, A. Ouerghi, O.E. Kusmartseva, F.V. Kusmartsev, M. Oueslati,{ \it Raman Spectroscopy of four epitaxial graphene layers: Macro-island grown on 4H-SiC (000) substrate and an associated strain distribution}, Thin Solid Films 538 (2013), doi: 10.1016/j.tsf.2013.05.093.
%[43] Raman Spectroscopy of four epitaxial graphene layers: Macro-island grown on 4H-SiC (000) substrate and an associated strain distribution
\bibitem{Das} A. Das, S. Pisana, B. Chakraborty, S. Piscanec, S. K. Saha, U. V. Waghmare, K. S. Novoselov, H. R. Krishnamurthy, A. K. Geim, A. C. Ferrari  and  A. K. Sood, {\it Monitoring Dopants by Raman Scattering in an Electrochemically Top-Gated Graphene Transistor}, Nature Nanotechnol. 3, (2008), pp. 210-215.  
%[44]Roll-to-roll production of 30-inch graphene films for transparent electrodes
\bibitem{Bae} S. Bae {\it et al.}, {\it  Roll-to-roll production of 30-inch graphene films for transparent electrodes}, Nature Nanotechnol. 5 (2010), pp. 574-578.
%[45]Intrinsic and extrinsic performance limits of graphene devices on SiO2
\bibitem{Chen3} J. H. Chen, C. Jang, S. Xiao, M. Ishigami and M. S. Fuhrer, {\it Intrinsic and extrinsic performance limits of graphene devices on $SiO_2$}, Nature Nanotechnol. 3 (2008), pp. 206-209.
%[46]Ultrahigh electron mobility in suspended graphene
\bibitem{Bolotin} K. I. Bolotin {\it et al.}, {\it Ultrahigh electron mobility in suspended graphene}, Solid State Commun. 146 (2008), pp. 351-355.
%[47]Approaching ballistic transport in suspended graphene
\bibitem{Du} X. Du, I. Skachko, A. Barker and E. Y. Andrei, {\it Approaching ballistic transport in suspended graphene}, Nature Nanotechnol. 3 (2008), pp. 491-495.
%[48]Giant intrinsic carrier mobilities in graphene and its bilayer
\bibitem{Morozov} S. V. Morozov {\it et al.} {\it Giant intrinsic carrier mobilities in graphene and its bilayer}, Phys. Rev. Lett. 100 (2008), p. 016602.
%[49]Semimetallic properties of a heterojunction
 \bibitem{Kusmartsev} F. V. Kusmartsev and A. M. Tsvelik, {\it Semi-metallic properties of a heterojunction}, JETP Lett. 42 (1985), pp. 257-260.
%[50]Origin of the relatively low transport mobility of graphene grown through chemical vapor deposition
\bibitem{Song} H. S. Song {\it et al.}, {\it Origin of the relatively low transport mobility of graphene grown through chemical vapor deposition}, Sci. Rep. (Nature) 2 (2012), p. 337.


%[51]Electronic transport in two-dimensional graphene
\bibitem{Sarma} S. D. Sarma, S. Adam, E. H. Hwang and E. Rossi, {\it Electronic transport in two-dimensional graphene}, Rev. Mod. Phys. 83 (2011), pp. 407-470. 
%[52] Magnetoresistance in graphene under quantum limit regime
\bibitem{Yang-Bo} Y. B. Zhou, H. C. Wu, D. P. Yu, and Z. M. Liao, {\it Magnetoresistance in graphene under quantum limit regime}, Appl. Phys. Lett. 102, (2013), p. 093116.
%[53]Electronic Transport in Mesoscopic SystemsElectronic Transport in Mesoscopic Systems
\bibitem{Datta} S. Datta, { \it Electronic Transport in Mesoscopic Systems}, Cambridge: Cambridge University Press (1997).
%[54]Direct nanoscale imaging of ballistic and diffusive thermal transport in graphene nano-structures
\bibitem{Pumarol} M. E. Pumarol {\it et al.}, {\it Direct nanoscale imaging of ballistic and diffusive thermal transport in graphene nano-structures}, Nano Lett. 12 (2012), pp. 2906-2911.
%[55]Graphene spreads the heat,
\bibitem{Prasher} R. Prasher, {\it Graphene spreads the heat}, Science 328 (2010), pp. 185-186.
%[56]Superior Thermal conductivity of Single-Layer Graphene
\bibitem{Balandin} A. A. Balandin {\it et al.}, {\it Superior Thermal conductivity of Single-Layer Graphene}, Nano Lett. 8 (2008), pp. 902-907.
%[57]Two-Dimensional Phonon Transport in Supported Graphene
\bibitem{Seol} J. H. Seol {\it et al.}, {\it  Two-Dimensional Phonon Transport in Supported Graphene}, Science 328 (2010), pp. 213-216.
%[58]Heat conduction across monolayer and few-layer graphenes
\bibitem{Koh} Y. K. Koh, M. H. Bae, D. G. Cahill and E. Pop, {\it Heat conduction across monolayer and few-layer graphenes}, Nano Lett. 10 (2010), pp. 4363-4368.
%[59]Fine structure constant defines transparency of graphene
\bibitem{Nair} R. R. Nair {\it et al.}, {\it  Fine structure constant defines transparency of graphene}, Science 320 (2008), pp. 1308-1308.
%[60]Graphene photonics and optoelectronics
\bibitem{Bonaccorso} F. Bonaccorso, Z. Sun, T. Hasan and A. C. Ferrari, {\it Graphene photonics and optoelectronics}, Nature Photonics  4 (2010), pp. 611-622.


%[61]Fluorescence of laser created electron-hole plasma in graphene
\bibitem{Stoehr} R. J. Stoehr, R. Kolesov, J. Pflaum, J. and J. Wrachtrup, {\it Fluorescence of laser created electron-hole plasma in graphene}, Phys. Rev. B  82 (2010), p. 121408.
%[62]Ultrafast carrier dynamics in graphite
\bibitem{Breusing} M. Breusing, C. Ropers and T. Elsaesser, {\it Ultrafast carrier dynamics in graphite}, Phys. Rev. Lett. 102 (2009), p. 086809.
%[63]Strongly coupled optical phonons in the ultrafast dynamics of the electronic energy and current relaxation in graphite 
\bibitem{Kampfrath} T. Kampfrath, L. Perfetti, F. Schapper, C. Frischkorn and M. Wolf, {\it Strongly coupled optical phonons in the ultrafast dynamics of the electronic energy and current relaxation in graphite}, Phys. Rev. Lett. 95 (2005), p. 187403.
%[64]Electronic transport and hot phonons in carbon nanotubes 
\bibitem{Lazzeri} M. Lazzeri, S. Piscanec, F. Mauri, A. C. Ferrari and J. Robertson, {\it Electronic transport and hot phonons in carbon nanotubes}, Phys. Rev. Lett. 95 (2005), p. 236802.
%[65]Stable forms of two-dimensional crystals and graphene
\bibitem{O'Hare2} V. V. Zalipaev, D. M. Forrester, C. M. Linton and F. V. Kusmartsev, {\it in the book New Progress on Graphene Research, Localised States of Fabry-Perot Type in Graphene Nano-Ribbons}, 
(http://www.intechopen.com/books/new-progress-on-graphene-research)
INTECH, (2013), Ch. 2.
%[66]Quantum Hall effect in graphene
\bibitem{Jiang} Z. Jiang, Y. Zhang, Y. W. Tana, H. L. Stormer and P. Kim, {\it Quantum Hall effect in graphene}, Solid State Comm. 143 (2007), pp. 14-19.
%[67]Chiral tunneling and the klein paradox in graphene
\bibitem{Katsnelson2} M. I. Katsnelson, K. S. Novoselov and A. K. Geim, {\it Chiral tunneling and the klein paradox in graphene}, Nature Phys. 2 (2006), pp. 620-625.
%[68]Paradox in a pencil
\bibitem{Calogeracos1} A. Calogeracos, {\it Paradox in a pencil}, Nature Phys. 2 (2006), pp. 579-580.
%[69]Barrier penetration and klein paradox
\bibitem{Su} R. K. Su, G. G. Siu and X. Chou, {\it Barrier penetration and klein paradox}, J. Phys A: Math. Gen. 26 (1993), pp. 1001-1005.
%[70]History and physics of klein paradox
\bibitem{Calogeracos2} A. Calogeracos and N. Dombey, {\it History and physics of klein paradox}, Contemp. Phys. 40 (1999), pp. 313-321.


%[71]Smooth electron waveguides in graphene
\bibitem{Hartmann}	R. R. Hartmann, N. J. Robinson, and M. E. Portnoi, {\it Smooth electron waveguides in graphene}. Phys. Rev. B, 81 (2010), pp. 245431.
%[72]Electronic fiber in graphene
\bibitem{Wu} Z. Wu, {\it Electronic fiber in graphene}, Appl. Phys. Lett. 98 (2011), pp. 082117.
%[73]Gate-controlled guiding of electrons in graphene
\bibitem{Williams} J. R. Williams, T. Low, M. S. Lundstrom, and  C. M. Marcus, {\it Gate-controlled guiding of electrons in graphene}, Nat. Nanotech. 6 (2011), pp. 222-225.  
%[74]Zero-energy states in graphene quantum dots and rings
\bibitem{Downing}C. A. Downing, D. A. Stone, and M. E. Portnoi, { \it Zero-energy states in graphene quantum dots and rings}, Phys. Rev. B 84  (2011), p. 155437. 
%[75]Searching for confined modes in graphene channels: The variable phase method
\bibitem{Stone}  D. A. Stone, C. A. Downing, M. E. Portnoi, { \it Searching for confined modes in graphene channels: The variable phase method}, Phys. Rev. B 86, (2012), p. 075464.
%[76]Fine-Structure and Fractional M/N Aharonov-Bohm Effect
\bibitem{Kusmartsev-1995} F.V. Kusmartsev, {\it Fine-Structure and Fractional M/N Aharonov-Bohm Effect},
Phys. Rev. B52 (1995), pp. 14445-14456.  
%[77] Interaction phenomena in graphene seen through quantum capacitance
\bibitem{Yu} G. L. Yu {\it et al.} {\it Interaction phenomena in graphene seen through quantum capacitance}, PNAS 110 (2013),  pp. 3282-3286.
%[78] Measurement of the quantum capacitance of graphene
\bibitem{Xia} J. Xia, F. Chen, J. Li and N. Tao, {\it  Measurement of the quantum capacitance of graphene}, Nature Nanotech. 4 (2009), pp. 505-509.
%[79] Screening length and quantum capacitance ing graphene by scanning probe microscopy
\bibitem{Giannazzo} F. Giannazzo, S. Sonde, V. Raineri and E. Rimini {\it  Screening length and quantum capacitance ing graphene by scanning probe microscopy}, Nano Lett. 9 (2009), pp. 23-29.
%[80] Density of states and its local fluctuations determined by capacitance of strongly disorder graphene
\bibitem{WLi} W. Li {\it et al.}, {\it Density of states and its local fluctuations determined by capacitance of strongly disorder graphene}, Scientific Rep. (Nature) 3 (2013), p. 1772.


%[81]Room-temperature transistor based on a single carbon nanotube
\bibitem{Tans} S. J. Tans, A. R. M. Verschueren and C. Dekker, {\it Room-temperature transistor based on a single carbon nanotube}, Nature 393 (1998), pp. 49-52.
%[82] Single- and multi-wall carbon nanotube field-effect transistors
\bibitem{Martel} R. Martel, T. Schmidt, H. R. Shea, T. Hertel and Ph. Avouris, {\it Single- and multi-wall carbon nanotube field-effect transistors}, Appl. Phys. Lett. 73 (1998), pp. 2447-2449.
%[83] Direct Synthesis of Long Single-Walled Carbon Nanotube Strands
\bibitem{Zhu} H. W. Zhu, C. L. Xu, D. H. Wu, B. Q. Wei, R. Vajtai, P. M. Ajayan, {\it  Direct Synthesis of Long Single-Walled Carbon Nanotube Strands}, Science 296 (2002), pp. 884-886.
%[84]Carbon-based electronics
\bibitem{Avouris} P. Avouris, Z. Chen and V. Perebeinos, {\it Carbon-based electronics}, Nature Nanotech. 2 (2007), pp. 605-615.
%[85]Symmetry of boundary conditions of the Dirac equation for electrons in carbon nanotubes
\bibitem{McCann} E. McCann and V. I. Fal'ko, {\it Symmetry of boundary conditions of the Dirac equation for electrons in carbon nanotubes}, Joural of Phys. Cond. Matter, 16 (2004), pp. 2371-2379.
%[86]Chemically derived, ultrasmooth graphene nanoribbon semiconductors 
\bibitem{Li2} X. Li, X. Wang, L. Zhang, S. Lee and H. Dai, {\it Chemically derived, ultra-smooth graphene nanoribbon semiconductors}, Science 319 (2008), pp. 1229-1232.
%[87]Synthesis of monolithic graphene-graphite integrated electronics
\bibitem{Park} J. Park, S. Nam, M. Lee and C. M. Lieber, {\it Synthesis of monolithic graphene-graphite integrated electronics}, Nature materials 98 (2011), p. 082117.
%[88] Graphene-based electronic sensors
\bibitem{He} Q. He, S. Wu, Z. Yin and H. Zhang, {\it Graphene-based electronic sensors}, Chem. Sci. 3 (2012), pp. 1764-1772.
%[89]Transparent, flexible, all-reduced graphene oxide thin film transistors
\bibitem{He2} Q. He {\it et al.}, {\it  Transparent, flexible, all-reduced graphene oxide thin film transistors}, ACS Nano 5 (2011), p. 082117.
%[90]Light-matter interaction in a microcavity-controlled graphene transistor
\bibitem{Engel} M. Engel {\it et al.}, {\it  Light matter interaction in a micro-cavity controlled graphene transistor room temperature transistor based on a single carbon nanotube}, Nature Commun. 3 (2012), pp. 906-911.


%[91] Extraordinary magnetoresistance: sensing the future
\bibitem{Tom} T. H. Hewett, F. V. Kusmartsev, {\it  Extraordinary magnetoresistance: sensing the future}, Central European Journal of Physics, 10 (2012), pp 602-608.
%[92] Element of Solid State Physics
\bibitem{Rudden} M. N. Rudden and J. Wilson, {\it Element of Solid State Physics}, New York Wiley (1993), Ch. 4-6.
%[93]A role for graphene in silicon-based semiconductor devices
\bibitem{Kim2} K. Kim, J. Y. Choi, T. Kim, S. H. Cho and H. J. Chung, {\it A role for graphene in silicon-based semiconductor devices}, Nature 479 (2011), pp. 338-344. 
%[94]Graphene double-layer capacitor with ac line-filtering performance
\bibitem{Miller} J. R. Miller, R. A. Outlaw and B. C. Holloway, {\it Graphene double-layer capacitor with ac line-filtering performance}, Science 329 (2010),  pp. 1637-1639.
%[95]Introduction to Electrical Engineering
\bibitem{Irwin} J. D. Irwin and D. V. Kerns, {\it Introduction to Electrical Engineering}, New Jersey: Prentice Hall, (1995), Ch. 8-9.
%[96]The Physics of Solids
\bibitem{Turton} R. Turton, {\it The Physics of Solids} New York: Oxford University Press, (2000), Ch. 4-6.
%[97] Charge neutrality and band-gap tuning of epitaxial graphene on SiC by molecular doping
\bibitem{Coletti} C. Coletti, C. Riedl, D. S. Lee, B. Krauss, L. Patthey, K. von Klitzing, J. H. Smet and U. Starke, {\it Charge neutrality and band-gap tuning of epitaxial graphene on SiC by molecular doping}, Phys. Rev. B 81 (2010), p. 235401.
%[98] The role of defects and doping 2D graphene sheets and 1D nanoribbons
\bibitem{Terrones} H. Terrones, R. Lv, M. Terrones and M. S. Dresselhaus, {\it The role of defects and doping 2D graphene sheets and 1D nanoribbons}, Rep. Prog. Phys. 75 (2012), p. 062501
%[90] Band gap opening of graphene by doping small boron nitride domains
\bibitem{Fan} X. Fan, Z. Shen, A. Q. Liu and J. L. Kuo, {\it Band gap opening of graphene by doping small boron nitride domains}, Nanoscale 4 (2012), pp. 2157-2165.
%[100] Direct band gap opening in graphene by BN doping: Ab initio calculations
\bibitem{Shinde} P. P. Shinde and V. Kumar, {\it Direct band gap opening in graphene by BN doping: Ab initio calculations}, Phys. Rev. B 84 (2011), p. 125401.


%[101] Modulation-doped growth of mosaic graphene with single-crystalline p-n junctions for efficient photocurrent generation
\bibitem{KYan} K. Yan, D. Wu, H. Peng, L. Jin, Q. Fu, X. Bao and Z. Liu, {\it  Modulation-doped growth of mosaic graphene with single-crystalline p-n junctions for efficient photocurrent generation}, Nature Comm. 3 (2012), p. 1280.
%[102] Electron-hole asymmetry and energy gaps in bilayer graphene
\bibitem{Kruczynski} M. M. Kruczynski, E. McCann and V. I. Fal'ko,  {\it Electron-hole asymmetry and energy gaps in bilayer graphene}, Semicond. Sci. Technol. 25 (2010),  p. 033001.
%[103]Conductance anomaly near the Lifshitz transition in strained bilayer graphene
\bibitem{Gradinar} D. Gradinar, H. Schomerus and V. Fal'ko, {\it Conductance anomaly near the Lifshitz transition in strained bilayer graphene}, Phys. Rev. B. 85 (2012), p. 165429.
%[104] Unraveling the acoustic electron-phonon interaction in graphene
\bibitem{Kaasbjerg} K. Kaasbjerg, K. S. Thygesen and K. W. Jacobsen, {\it Unraveling the acoustic electron-phonon interaction in graphene}, Phys. Rev. B 85 (2012), p. 165440.
%[105]Graphene oxide thin film field effect transistors without reduction
\bibitem{Jin} M. Jin {\it et al.}, {\it  Graphene oxide thin film field effect transistors without reduction}, J. Phys. D: Appl. Phys. 42 (2009), p. 135109.
%[106]Carbon nanotubes finally deliver
\bibitem{Kreupl} F. Kreupl, {\it Carbon nanotubes finally deliver}, Nature 484 (2012), pp. 321-322.
%[107]Sub-10 nm carbon nanotube transistor
\bibitem{Franklin} A. D. Franklin {\it et al.}, {\it Sub-10 nm carbon nanotube transistor}, Nano Lett. 12 (2012), pp. 758-762.
%[108]Flexible, transparent single-walled carbon nanotube transistors with graphene electrodes
\bibitem{Jang} Jang, S. {\it et al.} {\it Flexible, transparent single-walled carbon nanotube transistors with graphene electrodes},  Nanotech.  21 (2010), p. 425201. 
%[109]Coherent nonlocal transport in quantum wires with strongly coupled electrodes
\bibitem{Barbara} Y. Yang, G. Fedorov, P. Barbara, S. E. Shafranjuk, B. K. Cooper, R. M. Lewis and C. J. Lobb, {\it Coherent nonlocal transport in quantum wires with strongly coupled electrodes}, Physical Review B 87 (2013), p. 045403.
%[110]Direct observation of a widely tunable band-gap in bilayer graphene
\bibitem{Zhang3} Y. Zhang {\it et al.}, {\it Direct observation of a widely tunable band-gap in bilayer graphene}, Nature 459 (2009), pp. 820-823.


%[111]Tunable metal-insulator transition in double-layer graphene heterostructures
\bibitem{Ponomarenko} L. A. Ponomarenko {\it et al.}, {\it  Tunable metal-insulator transition in double-layer graphene heterostructures}, Nature Phys. 7 (2011), pp. 958-961.
%[112]Tunnel field-effect transistors as energy-efficient electronic switches
\bibitem{Ionescu} M. A. Ionescu and H. Reil, {\it Tunnel field-effect transistors as energy-efficient electronic switches}, Nature 479 (2011), pp. 329-337.
%[113] Common-path interference and oscillatory Zener tunneling in bilayer graphene p-n junctions
\bibitem{Nandkishore} R. Nandkishore and L. Levitov, {\it Common-path interference and oscillatory Zener tunneling in bilayer graphene p-n junctions}, PNAS 108 (2011), pp. 14021-14025.
%[114]]Model of tunneling transistors based on graphene on sic
\bibitem{Michetti} P. Michetti, M. Cheli and G. Iannaccone, {\it Model of tunneling transistors based on graphene on sic.}, Appl. Phys. Lett. 96 (2010), p. 133508.
%[115]Computational study of tunneling transistor based on graphene nanoribbon
\bibitem{Zhao} P. Zhao, J. Chauhan and J. Guo, {\it Computational study of tunneling transistor based on graphene nanoribbon}, Nano Lett. 9 (2009), pp. 684-688.
%[116]Graphene nanoribbon tunnel transistors
\bibitem{Zhang2} Q. Zhang, T. Fang, H. Xing and A. Seabaugh and D. Jena, {\it  Graphene nanoribbon tunnel transistors}, IEEE Electron Dev. Lett.  29 (2008), pp. 1344-1346.
%[117]Graphene tunneling fet and its applications in low power circuit design
\bibitem{Yang} X. Yang {\it et al.}, {\it Graphene tunneling FET and its applications in low power circuit design}, {\it GLSVLSI10 Proceedings of the 20th symposium on Great lakes symposium on VLSI} (2010), pp. 263-268.
%[118]Transport in graphene tunnel junctions
\bibitem{Malec} C. M. Malec and D. Davidovic, {\it Transport in graphene tunnel junctions}, J. Appl. Phys. 109 (2011), p. 064507.
%[119]Graphene as a tunnel barrier: graphene-based magnetic tunnel junctions
\bibitem{Cobas} E. Cobas, A. L. Friedman,O. M. J. Erve, J. T. Robinson and B. T. Jonker, {\it Graphene as a tunnel barrier: graphene-based magnetic tunnel junctions}, Nano Lett. 12 (2012), pp. 3000-3004. 
%[120]Field-effect tunneling transistor based on vertical graphene heterostructures,
\bibitem{Britnell} L. Britnell {\it et al.}, {\it Field-effect tunneling transistor based on vertical graphene heterostructures}, Science 335 (2012), pp. 947-950.


%[121] Vertical field-effect transistor based on graphene
\bibitem{Georgiou} T. Georgiou {\it et al.}, {\it  Vertical field-effect transistor based on graphene-WS2 heterostructures for flexible and transparent electronics}, Nature Nanotech.  8 (2013), pp. 100-103.
%[122]100-GHz transistors from wafer-scale epitaxial graphene
\bibitem{Lin} Y. M. Lin  {\it et al.}, {\it 100-GHz transistors from wafer-scale epitaxial graphene}, Science 327 (2010), p. 662.
%[123]High-frequency, scaled graphene transistors on diamond-like carbon
\bibitem{Wu2} Y. Wu {\it et al.}, {\it High-frequency, scaled graphene transistors on diamond-like carbon}, Nature 472 (2011), pp. 74-78. 
%[124]Sub-10 nm gate length graphene transistors: operating at terahertz frequencies with current saturation. 
\bibitem{Zheng} J. Zheng  {\it et al.}, {\it  Sub-10 nm gate length graphene transistors: operating at terahertz frequencies with current saturation}, Scientific Rep.  3 (2013), p. 1314. 
%[125]Current saturation in zero-band-gap, top-gated graphene field-effect transistors 
\bibitem{Meric} I. Meric {\it et al.}, {\it  Current saturation in zero-band-gap, top-gated graphene field-effect transistors}, Nature Nanotech. 3 (2008), pp. 654-659. 
%[126] A roadmap for graphene
\bibitem{Novoselov5} K. S. Novoselov, V. I. Fal'ko, L. Colombo, P. R. Gellert, M. G. Schwab and K. Kim, {\it  A roadmap for graphene}, Nature 490 (2012), pp. 192-200.
%[127]Strong, light, multifunctional fibers of carbon nanotubes with ultrahigh conductivity
\bibitem{Behabtu} N. Behabtu {\it et al.}, {\it  Strong, light, multifunctional fibers of carbon nanotubes with ultrahigh conductivity}, Science 339 (2013), pp. 182-186. 
%[128]Transistors arrive at the atomic limit,
\bibitem{Lansbergen} G. P. Lansbergen, {\it Transistors arrive at the atomic limit}, Nature Nanotech. 7 (2012), pp. 209-210.
%[129]Graphene nanoelectronics,
\bibitem{Westervelt} R. M. Westervelt, {\it Graphene nanoelectronics}, Science 320 (2008), pp. 324-325. 
%[130]Graphene single-electron transistors
\bibitem{Ihn} T. Ihn {\it et al.}, {\it  Graphene single-electron transistors}, Materialstoday 13 (2010), pp. 44-50.
%[131] Competition for Graphene: Graphynes with Direction-Dependent Dirac Cones
\bibitem{Malko} D. Malko, C. Neiss, F. Vines and A. Gorling, {\it  Competition for Graphene: Graphynes with Direction-Dependent Dirac Cones}, Phys. Rev. Lett. 108 (2012), p. 086804.
%[132]
\bibitem{Britnell2} L. Britnell {\it et al.}, {\it Resonant tunnelling and negative differential conductance in graphene transistors}, Nature Comm. 4:1974 (2013).



\end{thebibliography}
\end{document}